\documentclass[aps,prb,reprint,superscriptaddress,showpacs]{revtex4-1}
\usepackage{graphicx}
\usepackage{natbib}
\usepackage{amsmath}
\usepackage{amssymb}
\usepackage{amsthm}
\usepackage{bm}
\usepackage{amsmath}
\usepackage{amssymb}
\usepackage{amsthm}
\usepackage{graphicx}
\usepackage{empheq}
\usepackage{color}
\usepackage{float}
\usepackage{multirow}

\newcommand{\bra}[1]{\left<#1\right|} 
\newcommand{\ket}[1]{\left|#1\right>}

\begin{document}


\title{Hole-Spin-Echo Envelope Modulations}


\author{Pericles Philippopoulos}
\affiliation{Department of Physics, McGill University, 3600 rue University, Montreal, Qc H3A 2T8, Canada}

\author{Stefano Chesi}
\affiliation{Beijing Computational Science Research Center, Beijing 100193, China}
\affiliation{Department of Physics, Beijing Normal University, Beijing 100875, China}

\author{Joe Salfi}
\affiliation{School of Physics, The University of New South Wales, Sydney, New South Wales 2052, Australia}
\affiliation{Centre for Quantum Computation and Communication Technology, The University of New South Wales, Sydney, New South Wales 2052, Australia}
\affiliation{Department of Electrical and Computer Engineering, University of British Columbia, Vancouver, BC V6T 1Z4}

\author{Sven Rogge}
\affiliation{School of Physics, The University of New South Wales, Sydney, New South Wales 2052, Australia}
\affiliation{Centre for Quantum Computation and Communication Technology, The University of New South Wales, Sydney, New South Wales 2052, Australia}

\author{W. A. Coish}
\affiliation{Department of Physics, McGill University, 3600 rue University, Montreal, Qc H3A 2T8, Canada}


\date{\today}



\begin{abstract}
Hole spins in semiconductor quantum dots or bound to acceptor impurities show promise as potential qubits, partly because of their weak and anisotropic hyperfine couplings to proximal nuclear spins.  Since the hyperfine coupling is weak, it can be difficult to measure.  However, an anisotropic hyperfine coupling can give rise to a substantial spin-echo envelope modulation that can be Fourier-analyzed to accurately reveal the hyperfine tensor.  Here, we give a general theoretical analysis for hole-spin-echo envelope modulation (HSEEM), and apply this analysis to the specific case of a boron-acceptor hole spin in silicon.  For boron acceptor  spins in unstrained silicon, both the hyperfine and Zeeman Hamiltonians are approximately isotropic leading to negligible envelope modulations. In contrast, in strained silicon, where light-hole spin qubits can be energetically isolated, we find the hyperfine Hamiltonian and $g$-tensor are sufficiently anisotropic to give spin-echo-envelope modulations. We show that there is an optimal magnetic-field orientation that maximizes the visibility of envelope modulations in this case.  Based on microscopic estimates of the hyperfine coupling, we find that the maximum modulation depth can be substantial, reaching $\sim 10\%$, at a moderate laboratory magnetic field, $B\lesssim 200\,\mathrm{mT}$.     

\end{abstract}

\maketitle

\section{Introduction}\label{sec:Intro}

 
Recent theoretical\cite{salfi2016charge, abadillo2018entanglement} and experimental\cite{van2018readout,kobayashi2018} work has shown that hole spins bound to boron acceptors in silicon may be viable qubits.  In this  system, a strong spin-orbit coupling can be used to manipulate the spins with electric fields while the influence of electrical noise is suppressed.\cite{salfi2016charge} This type of all-electrical control is more difficult in electron-spin systems, where the spin-orbit interaction is weaker.  A further possible advantage of hole spins is a weak and anisotropic hyperfine coupling that can be controlled to extend spin coherence times.\cite{fischer2008spin}  A hole spin bound to an acceptor in silicon therefore offers certain important advantages over other spin qubits.

Hyperfine interactions can have a significant influence on the spin dynamics of both electrons and holes in semiconductor nanostructures.  To accurately control these spins, it is important to first experimentally extract details of the relevant hyperfine parameters.  Spectroscopic techniques (e.g., paramagnetic spin resonance) can be used to resolve the large hyperfine splittings for electrons bound to donor impurities.\cite{feher1959, fletcher1954}  However, these methods applied to, e.g., boron acceptors in silicon have not yet resolved the much smaller hyperfine couplings expected for $p$-like orbitals composing the valence band, for which the dominant contact interaction vanishes. For example, in Ref. \onlinecite{stegner2010} it has been estimated that the hyperfine field for boron acceptors must be less than $\sim 0.7\,\mathrm{mT}$ to be consistent with recent spin-resonance measurements.  This is in contrast with the larger hyperfine field of $A/g\mu_\mathrm{B}\simeq 3.6\,\mathrm{mT}$ experienced by a phosphorus-donor-bound electron in Si ($g\simeq 2$) due to the hyperfine coupling to a $^{31}\mathrm{P}$ nuclear spin ($A/h\simeq 100\,\mathrm{MHz}$). Spectral hole-burning experiments showing the transfer of spin polarization from boron acceptors to surrounding $^{29}\mathrm{Si}$ nuclear spins indicate that the hyperfine coupling must be finite, but of undetermined strength.\cite{dirksen1989esr} 

Although the hyperfine coupling has not yet been experimentally resolved for boron acceptors in pure silicon, electron-nuclear double resonance (ENDOR) experiments have established the hyperfine coupling and quadrupolar splittings for boron acceptors at several lattice sites and in several polymorphs of SiC (specifically, 3C-, 4H- and 6H-SiC).  These experiments give hyperfine fields on the order of $A/g\mu_\mathrm{B}\sim 0.1\,\mathrm{mT}$.\cite{muller1993endor,greulich1998epr} If the hyperfine coupling is similarly weak for boron acceptors in pure silicon, it may not have been visible in Ref.~\onlinecite{stegner2010}, but it would nevertheless have important implications for hole-spin dynamics.  

As an alternative to ENDOR, direct measurements of electron-spin-echo envelope modulation (ESEEM)\cite{rowan1965, mims1972envelope} can be a sensitive probe of the hyperfine coupling when that coupling is anisotropic.  This technique has been successfully applied, for example, to understand the anisotropic hyperfine coupling for weakly coupled $^{29}\mathrm{Si}$ nuclear spins surrounding a phosphorus donor impurity in isotopically enriched $^{29}\mathrm{Si}$\cite{abe2004} and for $^{13}\mathrm{C}$ nuclear spins weakly coupled to nitrogen-vacancy- (NV-) center spins in diamond.\cite{smeltzer2011}  The influence of echo envelope modulation on decoherence/dynamics for donor-bound electrons and NV-center spins has also been analyzed in detail theoretically.\cite{saikin2003,witzel2007,maze2008} 

Here, we theoretically establish conditions (e.g., strain, magnetic field) for an experiment to extract hyperfine parameters from \emph{hole}-spin-echo envelope modulations (HSEEM), applicable to an acceptor impurity or quantum-dot-bound hole spin.  In particular, we find that envelope modulations will be negligible for hole spins in unstrained silicon, but by introducing biaxial tensile strain, a light-hole spin qubit can show substantial modulations.  For concrete calculations, we focus on the case of a light hole bound to a boron acceptor in silicon, where we expect the effect of envelope modulations to be significant.  However, much of the analysis presented here translates naturally to hole spins at other acceptor sites or in semiconductor quantum dots in group IV or III-V materials having a valence band with states that transform according to the $\Gamma_8$ representation of the $T_d$ group  at the band extremum.  The HSEEM effect studied here has the same basic origin as ESEEM, introduced by Rowan, Hahn, and Mims (Refs.~\onlinecite{rowan1965, mims1972envelope}).  This effect is distinct from modulations arising from non-secular hyperfine couplings for heavy-hole spin qubits,\cite{wang2012,wang2015} or from measurement feedback effects.\cite{carter2014} 

There are several issues that distinguish the case of HSEEM from the more conventional ESEEM.  For example, in contrast with the case of a donor-bound electron, acceptor-bound hole spins have a highly anisotropic $g$-tensor.  This anisotropy, along with the anisotropy of the hyperfine interaction, result in a visibility (modulation depth) of envelope modulations that has a non-trivial dependence on the applied magnetic-field orientation (Fig.~\ref{fig:Vk}, below). The modulation depth also depends on the strength of the hyperfine interaction, relative to the nuclear-spin Larmor frequency. To establish the maximum experimentally achievable modulation depth, we have determined the optimal magnetic-field orientation for a boron acceptor in silicon.  In addition, we have estimated the form and typical energy scale determining the acceptor hyperfine tensor from a semiempirical microscopic analysis. 

From the estimated hyperfine tensor for a boron acceptor, we evaluate the echo envelope function for a light-hole spin qubit [Fig.~\ref{fig:Vk}(a), below]. We find substantial modulation amplitude ($\gtrsim 10\%$) in a moderately weak magnetic field ($B\lesssim 200 \,\mathrm{mT}$).  The maximum modulation depth can be achieved only when the magnetic-field orientation has been optimized [see Fig.~\ref{fig:Vk}(b)].  This suggests that (under reasonable, but carefully designed experimental conditions), the hyperfine tensor can be extracted for a boron acceptor spin.



The rest of this article is organized as follows: in Sec.~\ref{sec:HSEEM} we review HSEEM and explain how it can be used to measure hyperfine couplings. In Sec.~\ref{sec:estimateB} we estimate the hyperfine tensor for a boron acceptor in silicon.  In Sec.~\ref{sec:light-hole}, we describe how the hyperfine tensor from Sec.~\ref{sec:estimateB} can be combined with the general analysis of Sec.~\ref{sec:HSEEM} to predict envelope modulations for a light-hole spin qubit. In Sec.~\ref{sec:conclusion} we present our conclusions.



\section{Hole-spin-echo envelope modulations (HSEEM)}\label{sec:HSEEM}
\subsection{Spin Hamiltonians}\label{sec:SH}

The hyperfine interaction for a hole-spin qubit in contact with a nuclear spin $\mathbf{I}$ can generally be written (with $\hbar=1$) as\cite{slichter1978}
\begin{equation}\label{eq:Hhf1}
H_{\mathrm{hf}}=\mathbf{S} \cdot \overleftrightarrow{\mathbf{A}}\cdot\textbf{I}+\boldsymbol{{\cal{B}}}\cdot\mathbf{I},
\end{equation}
where $\mathbf{S}$ is a pseudospin-$1/2$ operator acting in the two-dimensional qubit Hilbert space, $\overleftrightarrow{\mathbf{A}}$ is the hyperfine tensor, and $\boldsymbol{{\cal{B}}}$ gives rise to a chemical shift that may depend on the hole wavefunction, but is generally independent of the value of the pseudospin. If the qubit under consideration is composed of a Kramers doublet (two states related by time reversal), $\boldsymbol{{\cal{B}}}$ vanishes identically.\cite{philippopoulos2019} In an applied magnetic field $\mathbf{B}$, the full Hamiltonian is
\begin{equation}\label{eq:Hfull}
H = \mu_\mathrm{B} \mathbf{B} \cdot \overleftrightarrow{\mathbf{g}} \cdot\mathbf{S} - \gamma \mathbf{B} \cdot \mathbf{I} + H_{\mathrm{hf}},
\end{equation}
where $\overleftrightarrow{\mathbf{g}}$ is the hole-spin $g$-tensor, $\mu_\mathrm{B}$ is the Bohr magneton, and $\gamma$ is the nuclear-spin gyromagnetic ratio. For simplicity, we take the direction of the magnetic field, $\mathbf{B}$, to define the $z$-axis ($\hat{\mathbf{z}}=\hat{\mathbf{B}}=\mathbf{B}/\left|\mathbf{B}\right|$). We neglect the quadrupolar interaction between the nuclear spin and an electric-field gradient (this becomes exact for a nuclear spin $I=1/2$ with a vanishing quadrupole moment or for a local cubic symmetry, leading to a vanishing electric-field gradient\cite{abragam1961principles}). The Hamiltonian, Eq.~\eqref{eq:Hfull}, can be rewritten in terms of the nuclear-spin Zeeman splitting and the hole pseudospin splitting due to an effective magnetic field, $\mathbf{B}_{\mathrm{eff}}=\mathbf{B}\cdot\overleftrightarrow{\mathbf{g}}$ as
\begin{equation}\label{eq:HSc}
H = \Omega_S S_c  - \omega_I I_z + H_{\mathrm{hf}},
\end{equation}
where $\Omega_S = \mu_\mathrm{B}\left|\mathbf{B}\cdot\overleftrightarrow{\mathbf{g}}\right|=\mu_\mathrm{B}\left|\mathbf{B}_{\mathrm{eff}}\right|$ is the pseudospin splitting due to $\mathbf{B}_{\mathrm{eff}}$, $\omega_I=\gamma B$ is the nuclear-spin Zeeman splitting, and $S_c=\hat{\mathbf{c}}\cdot\mathbf{S}$, where $\hat{\mathbf{c}}:=\hat{\mathbf{B}}_{\mathrm{eff}}=\mathbf{B}_{\mathrm{eff}}/B_{\mathrm{eff}}$.

When $\Omega_S\gg \left|A_{\alpha\beta}\right|$, we retain only the secular contributions (those that commute with $S_c$), giving 
\begin{eqnarray}
H & \simeq & H_0  =  \Omega_S S_c  - \omega_I I_z + H_{\mathrm{hf}}^0,\label{eq:Hquantaxes}\\
H^0_{\mathrm{hf}} & = & A_{cx}S_cI_x+A_{cy}S_cI_y+A_{cz}S_cI_z,\label{eq:Hhf2}
\end{eqnarray}
where we restrict to the case $\boldsymbol{{\cal{B}}}\rightarrow 0$. Terms that do not commute with the nuclear-spin Zeeman term, $\sim I_z$, are included since the nuclear-spin Zeeman energy may be comparable to the hyperfine parameters, $\omega_I\sim A_{\alpha\beta}$.  Equations \eqref{eq:Hquantaxes} and \eqref{eq:Hhf2} are the standard starting point for studies of electron-spin-echo envelope modulation.\cite{rowan1965,saikin2003,witzel2007}

\subsection{Hahn echo envelope}\label{sec:Hahn}
In a hole-spin Hahn echo experiment, the hole pseudospin is initially aligned with the effective magnetic field (i.e., along $\hat{\mathbf{c}}$). A $\pi/2$-pulse is then applied, resulting in an equal superposition of Zeeman eigenstates. After some time $\tau$, a $\pi$-pulse is performed to invert the spin, followed by a free evolution and detection after a further time $\tau$.  

The time-evolution operator describing the dynamics of this Hahn echo pulse sequence is given by \cite{rowan1965}

\begin{equation}\label{eq:Hahnecho}
U(2\tau)=U_0(\tau)R_{\pi}U_0(\tau)R_{\pi/2},
\end{equation} 
where 
\begin{equation}
U_0(t)=e^{-iH_0t}
\end{equation}
is the the time-evolution operator under the Hamiltonian $H_0$ [Eq.~\eqref{eq:Hquantaxes}] and 
\begin{equation}
R_{\theta}=e^{-iS_b\theta}
\end{equation}
represents a rotation by angle $\theta$ around the $\hat{\mathbf{b}}$ axis, with $\hat{\mathbf{a}}$, $\hat{\mathbf{b}}$, and $\hat{\mathbf{c}}$ forming a right-handed triad: $\hat{\mathbf{a}}=\hat{\mathbf{b}}\times\hat{\mathbf{c}}$.  We take $\pi$- and $\pi/2$-pulses to be instantaneous relative to the time scale of envelope modulations.

The echo envelope, $V(\tau)$, describes the coherence of this hole spin at the end of a Hahn echo sequence and is defined as
\begin{equation}\label{eq:echoEnv}
V(\tau) = 2 \mathrm{Tr}\{\rho(2\tau)\sigma_+\} = 2\left<\sigma_{+}(2\tau)\right>,
\end{equation}
where 
\begin{equation}
\sigma_{+} = S_a +i S_b,
\end{equation}
and the density operator at the end of the sequence is given by
\begin{equation}\label{eq:rhot}
\rho(2\tau)=U(2\tau)\rho(0)U^{\dagger}(2\tau),
\end{equation}
with $\rho(0)$ describing the initial state.

For explicit calculations, we now specialize to the case where the hole spin is prepared in the Zeeman ground state, $\ket{\Downarrow}\bra{\Downarrow}$, and the nuclear spin is described by a maximally mixed (infinite-temperature) state, so that 
\begin{equation}\label{eq:rho0}
\rho(0)=\frac{1}{2I+1}\mathbb{I}\otimes\ket{\Downarrow}\bra{\Downarrow},
\end{equation}
where $\ket{\Uparrow}\,\left(\ket{ \Downarrow}\right)$ is an eigenstate of $S_c$ with eigenvalue $+1/2\,(-1/2)$. The echo envelope, $V(\tau)$, results from a sum over rotations of the hole pseudospin arising from each nuclear-spin Zeeman eigenstate, labeled by $m_I$, the eigenvalue of $I_{z}$. A consequence of the choice of (infinite-temperature) initial conditions [Eq.~\eqref{eq:rho0}] is that the average $\left<S_b(2\tau)\right>=0$ is preserved throughout the evolution and $V(\tau)=2\left<S_a(2\tau)\right>$ (a real quantity).  For a non-equilibrium (or low-temperature thermal) state of the nuclear-spin system, $V(\tau)$ may generally become complex and the specific expressions given below will not be realized.  The echo envelope $V(\tau)$ would, however, be constructed from the same Fourier components, realized from the eigenvalues of $H_0$. In an ensemble, a finite-temperature pseudo-pure hole-spin initial state will result in the same dynamics described here, but with a reduction in $V(\tau)$ by a factor $\epsilon\simeq \Omega_S/k_\mathrm{B}T$ at high spin temperature (for $\Omega_S/k_\mathrm{B}T\ll 1$).  

\subsection{Extracting hyperfine parameters}\label{sec:Measuring}

\begin{figure}
\centering
\includegraphics[width=\columnwidth]{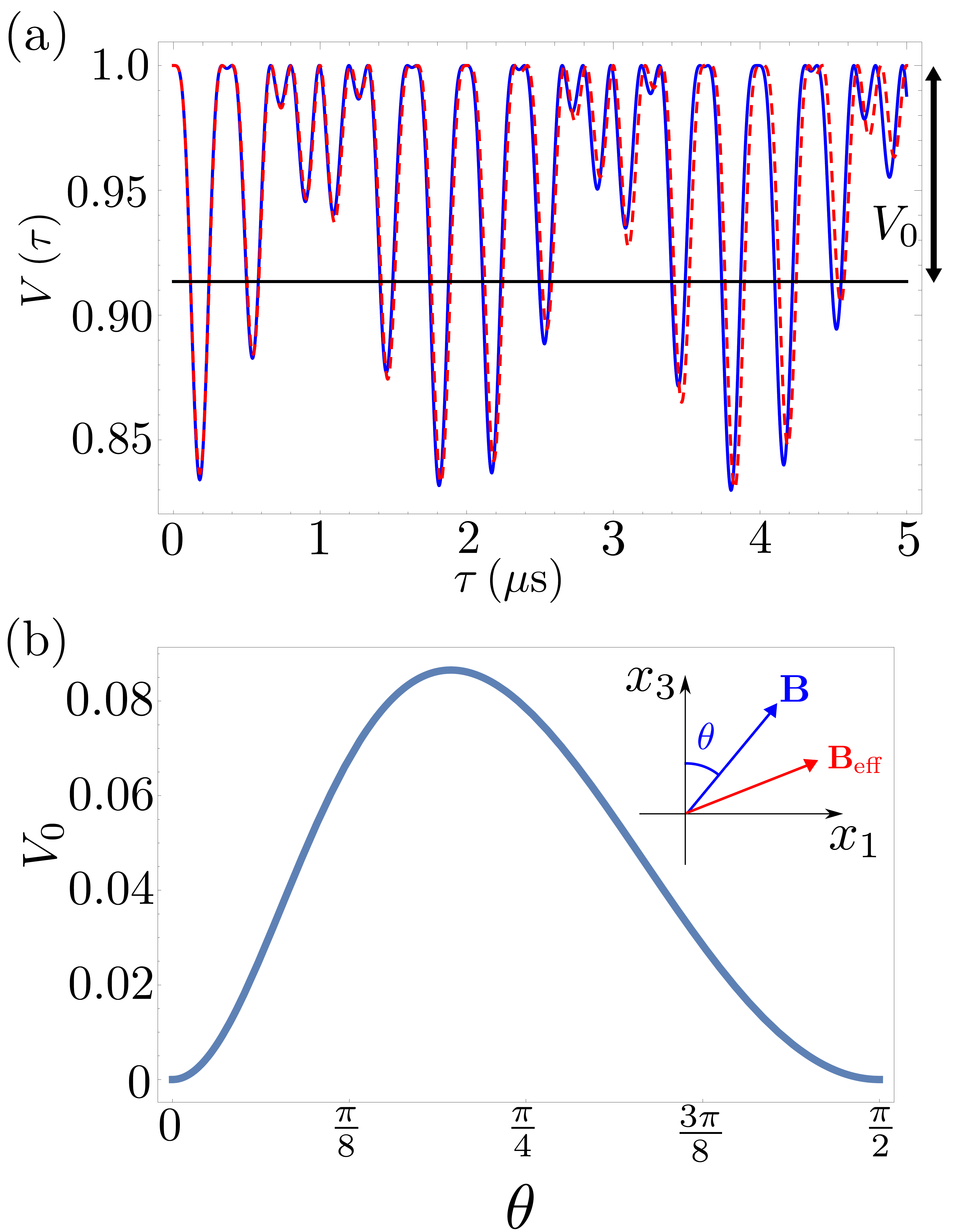}
\caption{(a) Spin-echo envelope function, $V(\tau)$, for a light hole bound to a $^{11}$B acceptor in silicon. The exact numerical result from Eq.~\eqref{eq:echoEnv} (blue solid line) is compared to the approximate form from Eq.~\eqref{eq:V} (red dashed line). The magnetic field is taken to have magnitude $B = 200\,\mathrm{mT}$ and is oriented to maximize the modulation depth, $\theta = \theta_{\mathrm{max}}\simeq 0.2\,\pi$ [see Fig.~\ref{fig:Vk}(b)]. (b) Modulation depth $V_0$ as a function of the magnetic field orientation angle $\theta$ (see inset) at a magnetic field strength of $B=200\,\mathrm{mT}$. The light holes transform like states of angular momentum $J_3 = \pm 1/2$ about the $x_3$ axis. For example, for biaxial tensile strain along $[100]$ and $[010]$, the ground-state doublet will be composed of $J_3 = \pm 1/2$ light-hole states where $x_3$ corresponds to the $[001]$ direction. The effective magnetic field, $\mathbf{B}_{\mathrm{eff}}$, is defined above Eq.~\eqref{eq:HSc} in the main text.  The modulation depth is independent of the azimuthal angle since both the Zeeman Hamiltonian [Eqs.~\eqref{eq:LHZeeman} and \eqref{eq:g}] and the hyperfine Hamiltonian, Eq.~\eqref{eq:LHHF}, are cylindrically symmetric in this case.}\label{fig:Vk}
\end{figure}

The full Hamiltonian (in the secular approximation and in a rotating frame at the hole Zeeman frequency with corresponding unitary $U_{Z}(t)=e^{i\Omega_S S_ct}$) is given, from Eqs.~\eqref{eq:Hquantaxes} and \eqref{eq:Hhf2}, by
\begin{equation}\label{eq:HsimWitzel}
\tilde{H}=\sum_{\beta} S_cA_{c\beta}I_{\beta} - \omega_I I_z,
\end{equation}
where $\beta \in \{x,y,z\}$. 

Because $H_{\mathrm{hf}}^0$ contains anisotropic terms ($A_{cx}S_cI_x$, $A_{cy}S_cI_y$), the nuclear-spin quantization axis depends on the state of the hole spin. Inverting the hole-spin orientation with a $\pi$-pulse during the Hahn echo sequence [Eq.~\eqref{eq:Hahnecho}] then results in an interference effect.  This produces a beating (modulations) in the echo envelope function.\cite{rowan1965} This signal can be used to extract information about the hyperfine constants $A_{c\beta}$.  There are well-known ways to do this extraction (see, e.g., Ref.~\onlinecite{abe2004} where the envelope-modulation frequencies were measured as a function of the magnetic-field orientation).  Here, for completeness, we illustrate how these parameters can be found in the present context and with a fixed magnetic-field orientation. For the Hamiltonian, $\tilde{H}$, and for the initial conditions given by Eq.~\eqref{eq:rho0}, the echo envelope can be approximated by\cite{mims1972envelope, mims1977nuclear} 
\begin{equation}\label{eq:V}
V(\tau) \approx 1-\frac{V_0}{2}[1-\cos(\omega_+\tau)][1-\cos(\omega_-\tau)],
\end{equation}
where 
\begin{equation}\label{eq:omegapm}
\omega_{\pm}=\sqrt{\left(\pm\frac{A_{cz}}{2}-\omega_I\right)^2+\left(\frac{A_{\mathrm{nc}}}{2}\right)^2}
\end{equation}
are the nuclear-spin precession frequencies, and where 
\begin{equation}\label{eq:Anc}
A_{\mathrm{nc}}=\sqrt{A_{cx}^2+A_{cy}^2}
\end{equation}
is the non-collinear part of the hyperfine interaction. In addition, the modulation depth $V_0$ is given by
\begin{equation}\label{eq:V0}
V_0=\frac{4}{3}I(I+1)k;\quad k=\frac{(\omega_I A_{\mathrm{nc}})^2}{(\omega_+\omega_-)^2},
\end{equation}
and $k$ is a parameter introduced in Ref.~\onlinecite{mims1972envelope} that gives the spin-echo modulation depth for a nuclear spin $I=1/2$. Equation \eqref{eq:V} applies in the limit of a small modulation depth, with corrections of order $\sim {\cal{O}}(k^2)$. \cite{mims1977nuclear} The parameter $k$ depends on both the orientation and the magnitude of the magnetic field, $\mathbf{B}$. The orientation of $\mathbf{B}$ determines the $\hat{\mathbf{c}}$ axis and $k$ depends on $\hat{\mathbf{c}}$ through $A_{\mathrm{nc}}$ [see Eqs.~\eqref{eq:Anc} and \eqref{eq:V0}]. We can therefore maximize $k$ with an appropriate choice for the magnetic-field orientation [Fig.~\ref{fig:Vk}(b)]. Furthermore, since $\omega_I \propto B$, if $\omega_I \gg A_{\mathrm{nc}}$, then $k\propto 1/B^2$ [see Eq.~\eqref{eq:V0}]. Thus, $k$ (and $V_0$) are strongly suppressed in a strong magnetic field.

From measurements of $V(\tau)$, the hyperfine parameters, $A_{cz}$ and $A_{\mathrm{nc}}$, can be determined independently via Fourier analysis, provided the nuclear-spin Larmor frequency, $\omega_I$, is known. If the qubit under consideration were not composed of a Kramers doublet, the nuclear-spin Larmor frequency may generally be affected by a finite chemical shift $\boldsymbol{{\cal{B}}}\ne 0$, introduced in Eq.~\eqref{eq:Hhf1}  (this could be the case, e.g., for a mixed heavy-hole/light-hole qubit system). In such a situation, $\omega_I$ should be measured independently to establish the hyperfine parameters. The analysis presented here has so far neglected a finite Hahn-echo decay time ($T_2$).  To extract the hyperfine coupling $\sim A_{cz}$ from envelope modulations, it is important that the envelope decay time be sufficiently long.  A minimal condition to resolve the coupling constant $A_{cz}$ is $T_2>\hbar/A_{cz}$.  We discuss this condition in the context of a light-hole qubit at a boron acceptor impurity in silicon in Sec.~\ref{sec:limitations}, below.

\section{Hyperfine tensor for a boron acceptor in silicon}\label{sec:estimateB}

Both the form and typical size of the hyperfine tensor can have an important influence on the visibility of envelope modulations.  To accurately estimate the hyperfine tensor, it is important to have a precise description of the electronic state in the immediate vicinity of the nucleus.  In general, this requires careful consideration of central-cell corrections, since the envelope-function approximation breaks down due to the $\sim 1/r$ singularity in the impurity potential.\cite{debernardi2006, usman2015, pica2014}   If we are willing to forgo a high degree of accuracy, a reasonable procedure is to start from the states evaluated within the envelope-function approximation and to adjust these states appropriately to account for short-ranged corrections.  A similar approach was followed by Kohn and Luttinger (Ref.~\onlinecite{kohn1955hyperfine}) in early work on the hyperfine coupling for a phosphorus donor impurity in silicon.  

Two corrections to the wavefunctions are required in the Kohn-Luttinger approach.  First, the bulk silicon Bloch functions should be replaced by atomic-like functions of appropriate symmetry and extent to characterize the wavefunction near the impurity nucleus.  Second, the electronic density should be rescaled to account for short-ranged corrections in the `central-cell' (this is equivalent to rescaling the envelope function very close to the impurity nucleus).  This central-cell correction is indicated by the fact that the binding energy calculated in the envelope-function approximation underestimates the true binding energy, suggesting a higher density near the impurity is required to account for the short-ranged potential.  To calculate the hyperfine coupling for a phosphorus donor, Kohn and Luttinger estimated a rescaling parameter based directly on the experimentally measured binding energy and the solution to an envelope-function equation in an exterior region (outside a specified cutoff radius).  This procedure has the disadvantage that the result depends on the choice of cutoff radius.  Here, we take a slightly different approach.  To estimate the rescaling parameter for a boron acceptor in silicon, we take advantage of the experimentally well-established hyperfine coupling for a phosphorus donor. The phosphorus donor and boron acceptor in silicon have similar binding energies (see Table \ref{tab:hf}, below), suggesting the two problems may have similar electrostatics.\footnote{Here, we use the term `electrostatics' to describe the problem of finding the full wavefunction of the impurity-bound hole.  This includes the long-range properties (band properties like the effective mass) and short-range properties (such as the core electronic structure of the impurity atom). If the electrostatics of the phosphorus donor and the boron acceptor are similar, then the rescaling parameter (central-cell corrections) should also be similar. While the similarity in binding energies is fully consistent with the two impurities having similar electrostatics, we cannot rule out the possibility of an accidental coincidence in binding energy. Ultimately, the true value of the hyperfine coupling (and central-cell correction) should be established experimentally, as proposed here.}  Based on this observation, we can extract the rescaling parameter from the known hyperfine coupling for a phosphorus donor and use it as an approximate proxy for the rescaling parameter of the boron acceptor.  This procedure avoids potential ambiguity in the choice of cutoff radius.  Based on the results obtained by Kohn and Luttinger for a donor impurity, we expect this procedure should give the strength of the hyperfine tensor for a boron acceptor to within a factor of $2$-$3$, sufficient to establish whether the tensor can be measured experimentally under reasonable conditions.         

For a donor or acceptor impurity, the wavefunction associated with spin $\sigma=\uparrow,\downarrow$ for a ground-state doublet labeled by pseudospin $\sigma^\prime$ can be approximated (far from the impurity nucleus) with the envelope-function approximation assuming $N_v$ degenerate valleys related by symmetry (e.g., $N_v = 6$ for the conduction band of silicon and $N_v = 1$ for the valence band of silicon):\cite{kohn1955theory}
\begin{equation}\label{eq:envfn}
\left<\mathbf{r}\sigma \right|\left.\sigma'\right>=\Psi^\sigma_{\sigma'}(\mathbf{r}) = \frac{1}{\sqrt{N_v}}\sum_{\nu\in \mathcal{S}_{\sigma'}}F_{\nu}(\mathbf{r})\psi_{\nu\mathbf{k}_\nu}^\sigma(\mathbf{r}).
\end{equation}
Here, $F_\nu(\mathbf{r})$ is an envelope function that solves the Schr\"odinger equation for a slowly varying impurity potential and with an effective-mass tensor associated with band/valley index $\nu$ at wavevector $\mathbf{k}=\mathbf{k}_{\nu}$. The symbol $\mathcal{S}_{\sigma'}$ indicates the subset of symmetry-related band/valley states associated with pseudospin $\sigma'$.  We have neglected spin-orbit coupling in the envelope equation, leading to a spin-independent envelope function $F_\nu(\mathbf{r})$. However, the Bloch functions $\psi_{\nu\mathbf{k}}^\sigma(\mathbf{r})$ solve the Schr\"odinger equation for the perfectly periodic bulk crystal potential including the short-ranged spin-orbit coupling:
\begin{equation}
\psi_{\nu\mathbf{k}}^\sigma(\mathbf{r}) = e^{i\mathbf{k}\cdot\mathbf{r}}u_{\nu\mathbf{k}}^\sigma(\mathbf{r}).
\end{equation}
The lattice-periodic Bloch amplitudes $u_{\nu\mathbf{k}}^\sigma(\mathbf{r})$ are normalized over the primitive-cell volume, $\Omega$, and the envelope functions are normalized over all space:
\begin{equation}\label{eq:normalization}
\sum_\sigma\int_{\Omega}d^3 r \left|u_{\nu\mathbf{k}}^\sigma(\mathbf{r})\right|^2=1;\quad\frac{1}{\Omega} \int d^3 r\left|F_\nu(\mathbf{r})\right|^2=1.
\end{equation}
To separate the isotropic and anisotropic hyperfine interactions, it is useful to expand the Bloch functions around the impurity site at $\mathbf{r}=0$ in terms of spherical harmonics,
\begin{equation}\label{eq:shexp}
\psi_{\nu\mathbf{k}_\nu}^\sigma(\mathbf{r}) = \sum_{lm}R_{lm}^{\sigma\nu}(r)Y_{l m}(\theta, \phi),
\end{equation}
where $R_{lm}^{\sigma\nu}(r)$ are radial functions. 

We first restrict to the case of donor states, where the dominant contribution to the hyperfine interaction arises from the $l=0$ ($s$-like) term in the expansion given in Eq.~\eqref{eq:shexp}. From the transformation properties of the valley states under crystalline and time-reversal symmetries, all $l=0$ contributions can be related to a common radial function $R_s(r)$, independent of $\sigma\nu$:
\begin{equation}
R^{\sigma \nu}_{00}(r)=\alpha_s R_s(r),\quad\nu\in\mathcal{S}_\sigma,\quad (\sigma=\sigma'),
\end{equation}
where $\alpha_s$ is a parameter controlling the degree of $s$-hybridization, and $R_s(r)$ describes the radial function in the immediate vicinity of the impurity, normalized to an atomic volume ($\frac{4}{3}\pi r_0^3=\Omega/2$):
\begin{equation}
\int_0^{r_0}dr r^2\left|R_s(r)\right|^2=1.
\end{equation}

For the $l=0$ ($s$-like) contribution, only the isotropic Fermi contact interaction contributes, yielding the effective spin Hamiltonian for the conduction-band states:\cite{coish2009nuclear, kohn1955hyperfine} 
\begin{equation}
H_{\mathrm{hf}}^{C} = A^{i} \mathbf{S}\cdot\mathbf{I},
\end{equation} 
where $A^i$ is the contact hyperfine interaction, proportional to the on-site electronic density, and $i$ labels the nuclear isotope.  Within the envelope-function approximation, we take $F_\nu(\mathbf{r})\simeq F_\nu(0)$ to be approximately constant in the vicinity of the nucleus.  Consistent with this limit, applying Eq.~\eqref{eq:envfn}, and the expansion, Eq.~\eqref{eq:shexp}, gives the hyperfine coupling to a phosphorus-donor nuclear spin ($i=\mathrm{^{31}P}$) in silicon: \cite{kohn1955hyperfine,philippopoulos2019}
\begin{equation}\label{eq:cHFconstant}
A^{^{31}\mathrm{P}} = \frac{ \mu_0}{3\pi}\mu_B\gamma_{^{31}\mathrm{P}} N_v \left|\alpha_{s}\right|^2\left|F_{\mathrm{P}}(0)\right|^2\left|R_{s}(0)\right|^2.
\end{equation}
Here, $\mu_0$ is the vacuum permeability and $N_v = 6$ for the conduction band of silicon.  We have approximated the envelope function for a phosphorus donor with the form for an isotropic effective mass, $F_\nu(\mathbf{r})\simeq F_P(\mathbf{r})$, independent of the band/valley index $\nu$ (consistent with Kohn and Luttinger, Ref.~\onlinecite{kohn1955hyperfine}).   
The weight factor, $\left|\alpha_{s}\right|^2$, has been estimated in Ref.~\onlinecite{ohkawa1979} for a $P$ donor (by diagonalizing the $\bf{k}\cdot\bf{p}$ matrix presented in Ref.~\onlinecite{cardona1966}), giving  $\left|\alpha_s\right|^2 \approx 0.38$.   

We now consider states relevant to acceptor impurities, with pure $p$ symmetry (states that only have an $l=1$ contribution to their spherical harmonic expansion).  The top of the valence band in bulk silicon is fourfold degenerate.  This degeneracy is generally broken into two Kramers doublets (at zero magnetic field) due to confinement and strain in the vicinity of an acceptor impurity. Here, we consider the pair of states $\nu=\sigma'$ that transform like the $J_3=\pm 1/2$ (light-hole) states of the $\Gamma_8$ representation of the $T_d$ double group (which we label with $\nu=\sigma'=\Uparrow$ and $\nu=\sigma'=\Downarrow$), where $J_3$ characterizes angular momentum about the $x_3$ axis. The effective hyperfine Hamiltonian then takes the anisotropic form [found by projecting Eq.~\eqref{eq:hfVB} in Appendix \ref{app:hfVB} onto the light-hole subspace]\cite{chekhovich2013element, philippopoulos2019} 
\begin{equation}\label{eq:LHHF}
H_{\mathrm{hf}}^{\mathrm{LH}}=\frac{1}{3}A_{\parallel}^i\left[S_3I_3+2\left(S_1I_1+S_2I_2\right)\right],
\end{equation}
where $A_{\parallel}^i$ is the hyperfine parameter for isotope $i$ and $S_j$ and $I_j$, with $j \in \{1,2,3\}$, are (pseudo)spin operators obeying the usual commutation relations: $\left[S_j,S_k\right]=i\epsilon_{jkl}S_l$, $\left[I_j,I_k\right]=i\epsilon_{jkl}I_l$. The coordinate system defined by $x_j\,(j=1,2,3)$ is generally determined by strain/confinement in the vicinity of the acceptor,\footnote{In the case of a flat (quasi-2D) unstrained quantum dot, the effective mass results in a heavy-hole ($J_3=\pm 3/2$) ground state.  For an acceptor impurity or quantum dot under biaxial in-plane tensile strain along $[100]$ and $[010]$, the light-hole ($J_3=\pm 1/2$) states may describe the ground state.  The $x_3$ axis is the out-of-plane direction (e.g., the growth axis for a quantum dot defined in a 2D hole gas at a heterostructure interface). For example, for a growth axis along $[001]$, we could take $x_3 = [001]$, $x_1 =[100]$, and $x_2=[010]$.} independent of the orientation of the applied magnetic field $\mathbf{B}$ ($\propto\hat{\mathbf{z}}$) and the effective field acting on the hole spin due to an anisotropic $g$-tensor, $\mathbf{B}_\mathrm{eff}$ ($\propto\hat{\mathbf{c}}$).  The relationship between these quantities is indicated in the inset of Fig.~\ref{fig:Vk}(b).   

Averaging the hyperfine Hamiltonian as in the case of a phosphorus donor, described above, but now for states describing a boron acceptor impurity in silicon, gives\cite{philippopoulos2019}   
\begin{equation}\label{eq:aHFconstant}
A_{\parallel}^i = \frac{4\mu_0}{5 \pi}\mu_B\gamma_{i} \left|F_{\mathrm{B}}(0)\right|^2 \int_0^{r_0} dr \left|R_p(r)\right|^2/r,
\end{equation}
where $F_\nu(\mathbf{r})=F_{\mathrm{B}}(\bm{r})$ is the boron-acceptor envelope function and $R_p(r)$ is the radial part of the Bloch function at the valence-band maximum in silicon ($\mathbf{k}=0$), normalized over an atomic volume.  As described above, we have assumed pure $p$-like states, $\alpha_p=1$.  Accounting for $p$-$d$ hybridization of the acceptor state would, in general, modify the anisotropy of the hyperfine tensor.\cite{philippopoulos2019, chekhovich2013element}  In writing Eq.~\eqref{eq:aHFconstant}, we have furthermore neglected long-ranged contributions to the hyperfine interaction between a nuclear spin and electron/hole spin density in a distant unit cell.\cite{yafet1961,fischer2008spin} The radial function $R_p(r)$ is related to the radial functions from Eq.~\eqref{eq:shexp} through
\begin{eqnarray}
 R_{10}^{\uparrow\Uparrow}(r)&=&R_{10}^{\downarrow\Downarrow}(r)=\sqrt{\frac{2}{3}}R_p(r), \\ R_{11}^{\downarrow\Uparrow}(r)&=&R_{1-1}^{\uparrow\Downarrow}(r)=\sqrt{\frac{1}{3}}R_p(r),
 \end{eqnarray} 
where all other $R^{\sigma\nu}_{lm}(r)$ vanish and the numerical factors are determined by Clebsch-Gordan coefficients.\cite{philippopoulos2019}

To account for deviations in the wavefunctions in the central-cell region, we now make two adjustments to the usual envelope-function approximation, as described above.  First, the radial functions $R_{s(p)}(r)$ associated with the silicon Bloch functions are replaced with hydrogenic orbitals having an effective core charge determined by Hartree-Fock theory for free boron and phosphorus atoms (taken from Ref.~\onlinecite{clementi1963}).  Specifically, we replace $R_s(r)$ with a $3s$ radial function with effective charge $Z_P = 5.64$ for the phosphorus donor and $R_p(r)$ is replaced with a $2p$ hydrogenic radial function with effective charge $Z_B = 2.42$ for the boron acceptor.  Second, the scaling parameters $|F_{P(B)}(0)|^2$ should be determined to account for the central-cell correction.  With the normalization given in Eq.~\eqref{eq:normalization}, $|F_{P(B)}(0)|^2$ corresponds to the probability to find the electron/hole in the central-cell region.  For a phosphorus donor impurity, Eq.~\eqref{eq:cHFconstant} can be used to extract $|F_{P}(0)|^2$ from the known value of the hyperfine coupling,\cite{feher1959} $A^\mathrm{^{31}P}/2\pi=117\,\mathrm{MHz}$:
\begin{equation}\label{eq:ccScaleDonor}
\left|F_P(0)\right|^2 = 0.014.
\end{equation}  

Equation \eqref{eq:ccScaleDonor} should be contrasted with the result from a direct application of the Kohn-Luttinger envelope function, $\left|F_P^\mathrm{KL}(0)\right|^2=\Omega/\pi a_l a_t^2=0.0019$ with $a_j=a_0\kappa m_0/m_j;\,j=l,t$.  Here, $a_0$ is the Bohr radius, $m_0$ is a free-electron mass, $\kappa=11.7$ is the dielectric constant for bulk silicon, and the longitudinal/transverse effective masses are $m_l=0.98,\,m_t=0.19$.  Empirically, the hyperfine coupling is therefore enhanced by a factor of $\simeq 7$ relative to the value expected within the envelope-function approximation.  Based on density functional calculations for bulk silicon,\cite{philippopoulos2019} we find that replacing the silicon Bloch functions with a $3s$ hydrogenic function for phosphorus leads to a further enhancement of the hyperfine coupling by a factor of $\simeq 3$.  The combination of these two effects, due to central-cell corrections and strain in the vicinity of the donor, lead to more than an order-of-magnitude increase in the hyperfine coupling relative to what would be expected from a na\"ive application of the envelope-function approximation.    
  
To approximate the hyperfine coupling for a boron acceptor from Eq.~\eqref{eq:aHFconstant}, we take
\begin{equation}\label{eq:FBequivFP}
\left|F_{\mathrm{B}}(0)\right|^2\approx \left|F_{\mathrm{P}}(0)\right|^2, 
\end{equation}
with $\left|F_P(0)\right|^2$ established empirically from Eq.~\eqref{eq:ccScaleDonor}.  Equation \eqref{eq:FBequivFP} is justified by the observation that the binding energies of phosphorus donors and boron acceptors are similar (see Table \ref{tab:hf}), and hence that the electrostatics and resulting central-cell corrections may therefore be similar.  Note that effective mass theory fails in the vicinity of the impurity potential, so it is not clear what (if any) further corrections could be made to Eq.~\eqref{eq:FBequivFP} to account for the different effective masses in the conduction and valence bands.  The ultimate accuracy of Eq.~\eqref{eq:FBequivFP} should be determined experimentally through direct measurements of the hyperfine coupling constants.  Here, we use this relation only to establish the plausibility of measuring envelope modulations for a boron-acceptor-bound light-hole spin qubit under realizable experimental conditions.  Using the assumption given in Eq.~\eqref{eq:FBequivFP} gives 
\begin{equation}\label{eq:BHFest}
A_{\parallel}^{^{11}\mathrm{B}}/2\pi \approx 1 \, \mathrm{MHz}.
\end{equation}
This result is also displayed in Table \ref{tab:hf}. This hyperfine coupling is sufficiently small that it may not have been resolved in the experiments of Ref.~\onlinecite{stegner2010}, but should be observable in the HSEEM experiments described above. Although it is not clear that the values should be the same (or even comparable), the estimate given in Eq.~\eqref{eq:BHFest} for a boron acceptor in silicon is within a factor of $\sim 3$ of the hyperfine coupling for boron acceptors in SiC measured with ENDOR.\cite{muller1993endor,greulich1998epr} While the hyperfine coupling for boron acceptors in SiC may not be a good measure of the hyperfine coupling for boron acceptors in silicon, the ENDOR experiments described in Refs.~\onlinecite{muller1993endor,greulich1998epr} provide evidence that even the weak (relative to electrons) hyperfine interaction for holes can have a measurable effect. 

The estimated value given in Eq.~\eqref{eq:BHFest} has been used to generate the plots in Fig.~\ref{fig:Vk}, where we have considered the case of $^{11}\mathrm{B}$ as an example. The hyperfine coupling for a boron acceptor in silicon (as estimated here) is roughly two orders of magnitude smaller than the coupling for a phosphorus donor.  Nevertheless, an HSEEM experiment would show echo envelope modulations with reasonable visibility $V_0\gtrsim 0.1$ at sufficiently low magnetic fields $B\lesssim 200\,\mathrm{mT}$.

\begin{table}
 \begin{ruledtabular}
\begin{tabular}{lccccc}
  \multirow{2}{*}{$i$} & \multirow{2}{*}{$I$} & \multirow{2}{*}{$\delta_i$} & \multirow{2}{*}{$E$ (meV)} & \multirow{2}{*}{$\gamma_i/2\pi$ ($ \frac{\mathrm{MHz}}{\mathrm{T}}$)} & Hyperfine\\
    &  &  &  &  & ($\mathrm{MHz}$)\\
   \hline
   $^{31}\mathrm{P}$ & $1/2$ & $100\%$ &  $44$ [\onlinecite{burstein1956}]& $17$ & $A^i/h = 117$\,[\onlinecite{feher1959}] \\
   $^{10}\mathrm{B}$ & $3$ & $20\%$ &  $45$ [\onlinecite{burstein1956}]& $0.46 $ & $A^i_{\parallel}/h \approx 0.3$ \\   
   $^{11}\mathrm{B}$ & $3/2$ & $80\%$ & $45$ [\onlinecite{burstein1956}]& $1.4$ & $ A^i_{\parallel}/h \approx 1$ \\ 
\end{tabular}\caption{The isotope ($i$), nuclear spin ($I$), natural isotopic abundance ($\delta_i$), measured binding energy ($E$) from Ref.~\onlinecite{burstein1956}, gyromagnetic ratio ($\gamma_i$), and hyperfine parameters for a phosphorus donor and a boron acceptor in silicon. The value for $A^{^{31}\mathrm{P}}$ was measured in Ref.~\onlinecite{feher1959} and $A^i_{\parallel}$ for boron acceptors [see Eq.~\eqref{eq:LHHF}] was estimated, as described in the main text.  }\label{tab:hf}
 \end{ruledtabular}
 \end{table}

\section{HSEEM for a hole-spin qubit}
\label{sec:light-hole}

A boron-acceptor-bound hole spin will generally couple to the nuclear spin associated with either of the stable isotopes, $^{10}$B ($I=3$) or $^{11}$B ($I=3/2$).  In addition, the hole spin may also couple to proximal $^{29}$Si ($I=1/2$, natural abundance $4.7\%$). In what follows, we will neglect coupling to nearby $^{29}$Si.  This is justified either by the weaker overlap of the hole envelope function with nuclear spins far from the central boron potential, or by considering acceptors in isotopically purified $^{28}$Si/$^{30}$Si (both isotopes having nuclear spin $I=0$). Furthermore, since both boron isotopes have a nuclear spin $I > 1/2$ and the boron acceptor in silicon has tetrahedral ($T_d$) symmetry, the nuclear quadrupolar interaction does not generally vanish. However, when the quadrupolar spitting is much smaller than the anisotropic part of the hyperfine interaction, the quadrupolar interaction can be neglected in calculating echo-envelope modulations.\cite{muller1993endor} We are unaware of direct measurements for the quadrupolar splittings or hyperfine couplings for boron acceptors in silicon.  However, ENDOR measurements on several polymorphs of SiC give quadrupolar splittings that are $ \sim 10\%$ of the measured value of $A_{\mathrm{nc}}$.\cite{muller1993endor,greulich1998epr}   The crystalline and electronic structure of SiC differs from that of silicon and may generally lead to distinct values of the hyperfine coupling and quadrupolar splitting.  Nevertheless, our estimated value of the hyperfine coupling for a boron acceptor in silicon is comparable to the measured values for boron acceptors in SiC (see Sec.~\ref{sec:estimateB}). Provided the quadrupolar splittings in both materials are also similar, ignoring the quadrupolar interaction is well justified here.  

In Ref.~\onlinecite{kobayashi2018}, hole-spin-echo experiments have been carried out for both strained and unstrained samples.  Without externally applied strain, the ground space of a boron acceptor in silicon is fourfold degenerate, spanned by the heavy-hole ($J_3 = \pm 3/2$) and light-hole states ($J_3 =\pm1/2$). We neglect the small anisotropy term, [$q\simeq 0$ in Eq.~\eqref{eq:ZeemanLutt}] resulting in a valence-band Zeeman Hamiltonian $\propto \mathbf{J}\cdot\mathbf{B}$ under an applied magnetic field $\mathbf{B}$.  We further consider pure $p$-like states ($\alpha_p=1$), giving $A_\perp^i=0$ in Eq.~\eqref{eq:HhfVBGeneral}, resulting in a hyperfine coupling $\propto \mathbf{J}\cdot\mathbf{I}$ [Eq.~\eqref{eq:hfVB}].  Under these conditions, the secular hyperfine coupling (the part that commutes with the valence-band Zeeman Hamiltonian) will commute with the nuclear Zeeman term for any orientation of the applied magnetic field, $\mathbf{B}$.  The result is that the nuclear-spin quantization axis will be independent of the hole-spin state, and there will be no echo envelope modulations under these conditions.  In contrast, under biaxial tensile strain, the ground space is spanned by only the light-hole ($J_3=\pm1/2$) states [see, for example, Ref.~\onlinecite{bir1963spin} or Eq.~(1) in the supplemental material of Ref.~\onlinecite{kobayashi2018}].  For light-hole states, both the effective Zeeman Hamiltonian [Eq.~\eqref{eq:LHZeeman} in Appendix \ref{app:g}] and the hyperfine Hamiltonian [Eq.~\eqref{eq:LHHF}] show strong anisotropy.  In this case, we expect substantial envelope modulations for the appropriate orientation of $\mathbf{B}$.  Up to small corrections, the dynamics of the acceptor-bound light-hole spin coupled to a boron nuclear spin will therefore be well described by the analysis presented in Sec.~\ref{sec:Measuring}.  

For a light-hole spin bound to a boron acceptor (boron isotope $i$) in silicon, $A^i_{cz}$ and $A^i_{\mathrm{nc}}$ can be determined using the light-hole hyperfine tensor [Eq.~\eqref{eq:LHHF}] and the $g$-tensor described in Appendix~\ref{app:g}, following the procedure outlined in Sec.~\ref{sec:SH}. This procedure gives
\begin{eqnarray}
A^i_{cz} & = & \frac{\sqrt{5-3\cos(2\theta)}}{3\sqrt{2}}A_{\parallel}^{i},\label{eq:AczTheta}\\
A^i_{\mathrm{nc}} & = & \frac{1}{2}\sqrt{\frac{1- \cos(4\theta)}{5 - 3\cos(2\theta)}}A_{\parallel}^{i}.\label{eq:AncTheta}
\end{eqnarray}
Inserting Eqs.~\eqref{eq:AczTheta} and \eqref{eq:AncTheta} into Eq.~\eqref{eq:V0} gives the orientation ($\theta$) dependence of the modulation depth, $V_0$. In Fig.~\ref{fig:Vk}(a), we show the echo envelope, $V(\tau)$, for a light-hole spin bound to a $^{11}\mathrm{B}$ acceptor. The magnetic field orientation $\theta$ has been chosen to maximize the modulation depth, $V_0$ [Fig.~\ref{fig:Vk}(b)]. The plots have been obtained using the estimate for $A_{\parallel}^{^{11}\mathrm{B}}$ discussed in Sec.~\ref{sec:estimateB}.  Recent spin-echo experiments have been performed on an ensemble of boron acceptors in strained silicon, where the light-hole spin can be energetically isolated.\cite{kobayashi2018} In these experiments, the applied magnetic field was oriented in-plane [corresponding to $\theta=\pi/2$ in Fig.~\ref{fig:Vk}(b)] to maintain a high quality factor for a superconducting resonator used for inductive detection.  Since we predict $V_0\simeq 0$ for $\theta=\pi/2$, we do not expect envelope modulations to be visible under the specific conditions of Ref.~\onlinecite{kobayashi2018}. However, our analysis predicts significant modulations for a similar experiment with an out-of-plane component of magnetic field.

For illustrative purposes, we have focused here on the case of a single boron acceptor in silicon, and have considered the example of $^{11}\mathrm{B}$. The same analysis gives the envelope modulation for a $^{10}\mathrm{B}$ acceptor if we account for the difference in nuclear spin $I$ and gyromagnetic ratio $\gamma_i$ for the two isotopes $i$ (see Table \ref{tab:hf}).  In particular, we can relate the two signals noting that both the nuclear Larmor frequencies and the hyperfine couplings are related by the gyromagnetic ratios: $\omega_I^i/\omega_I^{i'}=A_\parallel^i/A_\parallel^{i'}=\gamma_i/\gamma_{i'}$. This relationship assumes two isotopes of the same chemical species and neglects small corrections to the electronic structure due to the difference in nuclear mass.  The echo envelope signal arising from an ensemble of many boron acceptors including both isotopes will be given by a weighted average:
\begin{equation}
V(\tau) = \delta_{^{10}\mathrm{B}} V_{^{10}\mathrm{B}} (\tau) + \delta_{^{11}\mathrm{B}} V_{^{11}\mathrm{B}} (\tau),
\end{equation}
where $V_i(\tau)$ is the echo envelope for isotope $i$ [displayed in Eq.~\eqref{eq:V}], and $\delta_i$ is the isotopic abundance of isotope $i$. When the Larmor frequency of each isotope is much larger than the hyperfine coupling, $\omega^i_I \gg A^i_{\parallel}$, the modulation depth and frequencies determining $V_{^{10}\mathrm{B}} (\tau)$ are simply related to those for $V_{^{11}\mathrm{B}} (\tau)$ [see Eqs.~\eqref{eq:omegapm} and \eqref{eq:V0}]:
\begin{equation}\label{eq:rel}
\omega_{\pm}^{^{11}\mathrm{B}} = \frac{\gamma_{^{11}\mathrm{B}}}{\gamma_{^{10}\mathrm{B}}} \omega_{\pm}^{^{10}\mathrm{B}}; \quad V_0^{^{11}\mathrm{B}} \simeq \frac{ I(I+1)|_{I=\frac{3}{2}}}{ I(I+1)\rvert_{I=3}} V_0^{^{10}\mathrm{B}}, 
\end{equation}
where $\omega_{\pm}^i$ and $V_0^i$ are the frequencies and modulation depth parametrizing the echo envelope $V_i(\tau)$. If the isotopes in the sample are distributed according to the natural isotopic abundances (see Table~\ref{tab:hf}), Eq.~\eqref{eq:rel} results in modulation frequencies $\omega_{\pm}^{^{11}\mathrm{B}} \approx 3 \omega_{\pm}^{^{10}\mathrm{B}}$ and a relative amplitude $\delta_{^{11}\mathrm{B}}V_0^{^{11}\mathrm{B}}/(\delta_{^{10}\mathrm{B}}V_0^{^{10}\mathrm{B}}) \simeq \frac{5}{4}$.  Thus, for a natural isotopic abundance, we expect the envelope modulations arising from the two isotopes to have comparable amplitudes and the modulation frequencies are simply related by the gyromagnetic ratios of the two isotopes.


\subsection{Model limitations}\label{sec:limitations} 

Several limitations of the model have alredy been discussed.  Here, we collect and discuss the most significant limitations to be considered if an experiment is to accurately extract the hyperfine parameters.

\emph{Finite coherence time}---To extract the hyperfine couplings experimentally from envelope modulations, the Hahn-echo decay time, $T_2$, should exceed the typical inverse hyperfine coupling strengths.  From our estimate of the hyperfine coupling ($A_\mathrm{nc}/h\sim A_{cz}/h\sim A_\parallel/h\simeq 1\,\mathrm{MHz}$ for $^{11}$B as given in Table \ref{tab:hf}), this gives the requirement $T_2\gg \hbar/A_\parallel\simeq 0.2\,\mu\mathrm{s}$.  Recent experiments (Ref.~\onlinecite{kobayashi2018}) have demonstrated significantly longer Hahn-echo decay times, $T_2=0.9\,\mathrm{ms}$, at $B\simeq 200\,\mathrm{mT}$ for light-hole spins at boron acceptors in isotopically purified $^{28}\mathrm{Si}$. This suggests that recently measured coherence times are long enough to extract the hyperfine parameters from envelope modulations.  

\emph{Non-uniform strain distribution}---For an ensemble of acceptor impurities, a non-uniform strain distribution will generally lead to damped envelope modulations, on a time scale $t_\mathrm{damp}$.  This effect is commonly observed for phosphorus donor impurities in silicon, even without intentionally introducing additional strain.\cite{witzel2007}  For an experiment in which biaxial tensile strain is induced, as in Ref.~\onlinecite{kobayashi2018}, we estimate the variation in the hyperfine parameters $\delta A_{\alpha\beta}$ across the sample from the degree of heavy-hole/light-hole mixing generated by the Bir-Pikus Hamiltonian.\cite{bir1963spin, bir1963spinII} This gives a variation $\delta A_{\alpha\beta}/A_\parallel\sim\delta\epsilon/\epsilon$, where $\epsilon$ is the average strain along $x_1$ and $x_2$ and $\delta\epsilon$ describes the variation in strain accross the sample.  To resolve the modulation frequencies with a resolution $\lesssim A_\parallel$, we therefore require a sufficiently long damping time $t_\mathrm{damp}\sim\hbar/\delta A_{\alpha\beta}\gg\hbar/A_\parallel$.  In terms of strain, this condition requires a variation that is small compared to the average, $\delta \epsilon/\epsilon\ll 1$.

\section{Conclusions}\label{sec:conclusion}

Understanding the hyperfine coupling of valence-band (hole) spin states is an important step to predicting and controlling spin qubits derived from these states.  Boron-acceptor-spin qubits show potential for rapid local electric-field control, with relative immunity to electric-field noise,\cite{salfi2016charge, abadillo2018entanglement,kobayashi2018} but the hyperfine couplings for these qubits are still relatively poorly understood.  This hyperfine coupling may be small enough that it is difficult to resolve with certain spectroscopic techniques,\cite{stegner2010} but, as we have shown, it may lead to a significant echo envelope modulation at moderate magnetic fields.

As a concrete example, we have calculated the echo envelope function for a hole spin confined to a boron acceptor in silicon.  Without induced strain, the hole spin will show virtually no envelope modulations.  In contrast, a qubit defined in, e.g., the two-dimensional light-hole subspace will show substantial modulations.  This qubit can be energetically isolated through biaxial tensile strain.  The form of the hyperfine tensor in this case is given by the symmetry of the underlying electronic states.  By accounting for semiempirical corrections to the envelope-function approximation, we have estimated the typical size of the hyperfine parameter.  With this knowledge, we have shown that it is possible to maximize the visibility of envelope modulations with an appropriate orientation of the applied magnetic field.  Similarly, an experiment with the `incorrect' magnetic-field orientation [$\theta=0,\pi/2$ in Fig.~\ref{fig:Vk}(b)] would show no modulation at all. 

When the envelope modulations are visible, they can be used to accurately quantitatively determine the hyperfine tensor for a hole-spin qubit at an acceptor impurity, allowing for better control of hole-spin qubits and the nuclear spins that couple to them. 

There will be small corrections to the analysis presented here due, e.g., to the electric quadrupolar interaction with a nuclear spin $I>1/2$, $p$-$d$ hybridization of the microscopic electronic states, and further central-cell corrections that are not captured in our simple semiempirical approach.  Many of these effects (beyond the scope of the present work) could potentially be captured using \emph{ab initio} methods that have previously been applied to phosphorus donors in silicon.\cite{overhof2004ab}  An especially intriguing future question is whether the hyperfine tensor for an acceptor-spin qubit can be efficiently tuned or modulated through local strain or electric fields.  The strong anisotropy present in hole-spin hyperfine coupling may also provide an advantage in directly controlling the nuclear spin for a quantum memory or ancilla qubit.\cite{hodges2008}

\acknowledgments

WAC and PP acknowledge support from NSERC, CIFAR, FRQNT, and the Gordon Godfrey Bequest. SC acknowledges support from the National Key Research and Development Program of China (Grant No. 2016YFA0301200), NSAF (Grant No. U1530401) and NSFC (Grants No. 11574025 and No. 11750110428). JS acknowledges financial support from NSERC and from an ARC DECRA fellowship (DE160101490). SR acknowledges support from the ARC Centre of Excellence for Quantum Computation and Communication Technology (CE170100012), ARC Discovery Project (DP150103699), and in part from the U.S. Army Research Office (W911NF-17-1-0202). 

\appendix

\section{$g$-tensor for light holes}\label{app:g}
To obtain the $g$-tensor for light holes, we consider the Luttinger Hamiltonian. \cite{luttinger1956} The Zeeman term in the valence-band subspace of silicon can be written as\cite{luttinger1956,winkler2003}\footnote{The convention used here is that the pseudo-angular-momentum $\bf{J}$ measures the pseudo-angular-momentum of the crystal. Therefore, a hole in state $J_3 = m$ indicates that an electron with $J_3 = -m$ has been anihilated from the filled valence band, and the remaining member of the Kramers doublet has $J_3 = m$.}
\begin{equation}\label{eq:ZeemanLutt}
H_z=2\mu_B\kappa\mathbf{B}\cdot\mathbf{J} + 2\mu_Bq\mathbf{B}\cdot\boldsymbol{{\cal{J}}},
\end{equation}
where $\boldsymbol{{\cal{J}}}=(J_x^3,J_y^3,J_z^3)$. In the case of silicon, $\kappa=-0.42$ and $q=0.01$. \cite{winkler2003}  Since $|q| \ll |\kappa|$, we neglect the term proportional to $q$ for simplicity. Neglecting mixing with heavy-hole states, we project $H_z$ onto the light-hole subspace, and write the resulting matrix as 
\begin{equation}\label{eq:LHZeeman}
H_z^{\mathrm{LH}}=\mu_B \mathbf{B} \cdot \overleftrightarrow{\mathbf{g}} \cdot\mathbf{S},
\end{equation}
where
\begin{gather}\label{eq:g}
 \overleftrightarrow{\mathbf{g}}
 =
  2
  \begin{bmatrix}
  2\kappa & 0 & 0 \\
    0 & 2\kappa & 0 \\
    0 & 0 & \kappa
   \end{bmatrix}
\end{gather}
is the $g$-tensor for light holes in silicon. We note that the $g$-tensor is written in a basis $(x_1,x_2,x_3)$, where $x_3$ is the hole-spin quantization axis. 

\section{Hyperfine Hamiltonian in the valence band of silicon}\label{app:hfVB}

The hyperfine Hamiltonian projected onto the four-dimensional subspace spanned by the valence-band states at the $\Gamma$ point of silicon can be written as\cite{philippopoulos2019}
\begin{equation}\label{eq:HhfVBGeneral}
H^{\mathrm{VB}}_{\mathrm{hf}}=\left(\frac{1}{3}A^{i}_{\parallel}-\frac{3}{2}A^{i}_{\perp}\right) \textbf{J}\cdot \textbf{I} + \frac{2}{3}A^{i}_{\perp} {\cal{J}}\cdot \bf{I},
\end{equation}
where $A^{i}_{\parallel}$ and $A^{i}_{\perp}$ are hyperfine parameters. We neglect $A^{i}_{\perp}$ in our calculations, consistent with valence-band states that are pure $p$-states ($\alpha_p=1$) leaving
\begin{equation}\label{eq:hfVB}
H^{\mathrm{VB}}_{\mathrm{hf}}\approx \frac{1}{3}A^{i}_{\parallel} \textbf{J}\cdot \textbf{I}.
\end{equation}
Thus, in the limit of pure $p$-states, the hyperfine interaction is invariant under simultaneous rotations of the pseudospin-$3/2$, $\mathbf{J}$, and the nuclear spin, $\mathbf{I}$.  



\bibliography{HSEEM_Boron}

\begin{thebibliography}{46}%
\makeatletter
\providecommand \@ifxundefined [1]{%
 \@ifx{#1\undefined}
}%
\providecommand \@ifnum [1]{%
 \ifnum #1\expandafter \@firstoftwo
 \else \expandafter \@secondoftwo
 \fi
}%
\providecommand \@ifx [1]{%
 \ifx #1\expandafter \@firstoftwo
 \else \expandafter \@secondoftwo
 \fi
}%
\providecommand \natexlab [1]{#1}%
\providecommand \enquote  [1]{``#1''}%
\providecommand \bibnamefont  [1]{#1}%
\providecommand \bibfnamefont [1]{#1}%
\providecommand \citenamefont [1]{#1}%
\providecommand \href@noop [0]{\@secondoftwo}%
\providecommand \href [0]{\begingroup \@sanitize@url \@href}%
\providecommand \@href[1]{\@@startlink{#1}\@@href}%
\providecommand \@@href[1]{\endgroup#1\@@endlink}%
\providecommand \@sanitize@url [0]{\catcode `\\12\catcode `\$12\catcode
  `\&12\catcode `\#12\catcode `\^12\catcode `\_12\catcode `\%12\relax}%
\providecommand \@@startlink[1]{}%
\providecommand \@@endlink[0]{}%
\providecommand \url  [0]{\begingroup\@sanitize@url \@url }%
\providecommand \@url [1]{\endgroup\@href {#1}{\urlprefix }}%
\providecommand \urlprefix  [0]{URL }%
\providecommand \Eprint [0]{\href }%
\providecommand \doibase [0]{http://dx.doi.org/}%
\providecommand \selectlanguage [0]{\@gobble}%
\providecommand \bibinfo  [0]{\@secondoftwo}%
\providecommand \bibfield  [0]{\@secondoftwo}%
\providecommand \translation [1]{[#1]}%
\providecommand \BibitemOpen [0]{}%
\providecommand \bibitemStop [0]{}%
\providecommand \bibitemNoStop [0]{.\EOS\space}%
\providecommand \EOS [0]{\spacefactor3000\relax}%
\providecommand \BibitemShut  [1]{\csname bibitem#1\endcsname}%
\let\auto@bib@innerbib\@empty
\bibitem [{\citenamefont {Salfi}\ \emph {et~al.}(2016)\citenamefont {Salfi},
  \citenamefont {Mol}, \citenamefont {Culcer},\ and\ \citenamefont
  {Rogge}}]{salfi2016charge}%
  \BibitemOpen
  \bibfield  {author} {\bibinfo {author} {\bibfnamefont {J.}~\bibnamefont
  {Salfi}}, \bibinfo {author} {\bibfnamefont {J.~A.}\ \bibnamefont {Mol}},
  \bibinfo {author} {\bibfnamefont {D.}~\bibnamefont {Culcer}}, \ and\ \bibinfo
  {author} {\bibfnamefont {S.}~\bibnamefont {Rogge}},\ }\href@noop {}
  {\bibfield  {journal} {\bibinfo  {journal} {\prl}\ }\textbf {\bibinfo
  {volume} {116}},\ \bibinfo {pages} {246801} (\bibinfo {year}
  {2016})}\BibitemShut {NoStop}%
\bibitem [{\citenamefont {Abadillo-Uriel}\ \emph {et~al.}(2018)\citenamefont
  {Abadillo-Uriel}, \citenamefont {Salfi}, \citenamefont {Hu}, \citenamefont
  {Rogge}, \citenamefont {Calder{\'o}n},\ and\ \citenamefont
  {Culcer}}]{abadillo2018entanglement}%
  \BibitemOpen
  \bibfield  {author} {\bibinfo {author} {\bibfnamefont {J.~C.}\ \bibnamefont
  {Abadillo-Uriel}}, \bibinfo {author} {\bibfnamefont {J.}~\bibnamefont
  {Salfi}}, \bibinfo {author} {\bibfnamefont {X.}~\bibnamefont {Hu}}, \bibinfo
  {author} {\bibfnamefont {S.}~\bibnamefont {Rogge}}, \bibinfo {author}
  {\bibfnamefont {M.~J.}\ \bibnamefont {Calder{\'o}n}}, \ and\ \bibinfo
  {author} {\bibfnamefont {D.}~\bibnamefont {Culcer}},\ }\href@noop {}
  {\bibfield  {journal} {\bibinfo  {journal} {Appl.~Phys.~Lett.}\ }\textbf
  {\bibinfo {volume} {113}},\ \bibinfo {pages} {012102} (\bibinfo {year}
  {2018})}\BibitemShut {NoStop}%
\bibitem [{\citenamefont {van~der Heijden}\ \emph {et~al.}(2018)\citenamefont
  {van~der Heijden}, \citenamefont {Kobayashi}, \citenamefont {House},
  \citenamefont {Salfi}, \citenamefont {Barraud}, \citenamefont
  {Lavi{\'e}ville}, \citenamefont {Simmons},\ and\ \citenamefont
  {Rogge}}]{van2018readout}%
  \BibitemOpen
  \bibfield  {author} {\bibinfo {author} {\bibfnamefont {J.}~\bibnamefont
  {van~der Heijden}}, \bibinfo {author} {\bibfnamefont {T.}~\bibnamefont
  {Kobayashi}}, \bibinfo {author} {\bibfnamefont {M.~G.}\ \bibnamefont
  {House}}, \bibinfo {author} {\bibfnamefont {J.}~\bibnamefont {Salfi}},
  \bibinfo {author} {\bibfnamefont {S.}~\bibnamefont {Barraud}}, \bibinfo
  {author} {\bibfnamefont {R.}~\bibnamefont {Lavi{\'e}ville}}, \bibinfo
  {author} {\bibfnamefont {M.~Y.}\ \bibnamefont {Simmons}}, \ and\ \bibinfo
  {author} {\bibfnamefont {S.}~\bibnamefont {Rogge}},\ }\href@noop {}
  {\bibfield  {journal} {\bibinfo  {journal} {Sci.~Adv.}\ }\textbf {\bibinfo
  {volume} {4}},\ \bibinfo {pages} {eaat9199} (\bibinfo {year}
  {2018})}\BibitemShut {NoStop}%
\bibitem [{\citenamefont {Kobayashi}\ \emph {et~al.}(2018)\citenamefont
  {Kobayashi}, \citenamefont {Salfi}, \citenamefont {van~der Heijden},
  \citenamefont {Chua}, \citenamefont {House}, \citenamefont {Culcer},
  \citenamefont {Hutchison}, \citenamefont {Johnson}, \citenamefont {McCallum},
  \citenamefont {Riemann} \emph {et~al.}}]{kobayashi2018}%
  \BibitemOpen
  \bibfield  {author} {\bibinfo {author} {\bibfnamefont {T.}~\bibnamefont
  {Kobayashi}}, \bibinfo {author} {\bibfnamefont {J.}~\bibnamefont {Salfi}},
  \bibinfo {author} {\bibfnamefont {J.}~\bibnamefont {van~der Heijden}},
  \bibinfo {author} {\bibfnamefont {C.}~\bibnamefont {Chua}}, \bibinfo {author}
  {\bibfnamefont {M.~G.}\ \bibnamefont {House}}, \bibinfo {author}
  {\bibfnamefont {D.}~\bibnamefont {Culcer}}, \bibinfo {author} {\bibfnamefont
  {W.~D.}\ \bibnamefont {Hutchison}}, \bibinfo {author} {\bibfnamefont {B.~C.}\
  \bibnamefont {Johnson}}, \bibinfo {author} {\bibfnamefont {J.~C.}\
  \bibnamefont {McCallum}}, \bibinfo {author} {\bibfnamefont {H.}~\bibnamefont
  {Riemann}},  \emph {et~al.},\ }\href@noop {} {\bibfield  {journal} {\bibinfo
  {journal} {arXiv:1809.10859}\ } (\bibinfo {year} {2018})}\BibitemShut
  {NoStop}%
\bibitem [{\citenamefont {Fischer}\ \emph {et~al.}(2008)\citenamefont
  {Fischer}, \citenamefont {Coish}, \citenamefont {Bulaev},\ and\ \citenamefont
  {Loss}}]{fischer2008spin}%
  \BibitemOpen
  \bibfield  {author} {\bibinfo {author} {\bibfnamefont {J.}~\bibnamefont
  {Fischer}}, \bibinfo {author} {\bibfnamefont {W.~A.}\ \bibnamefont {Coish}},
  \bibinfo {author} {\bibfnamefont {D.~V.}\ \bibnamefont {Bulaev}}, \ and\
  \bibinfo {author} {\bibfnamefont {D.}~\bibnamefont {Loss}},\ }\href@noop {}
  {\bibfield  {journal} {\bibinfo  {journal} {\prb}\ }\textbf {\bibinfo
  {volume} {78}},\ \bibinfo {pages} {155329} (\bibinfo {year}
  {2008})}\BibitemShut {NoStop}%
\bibitem [{\citenamefont {Feher}(1959)}]{feher1959}%
  \BibitemOpen
  \bibfield  {author} {\bibinfo {author} {\bibfnamefont {G.}~\bibnamefont
  {Feher}},\ }\href@noop {} {\bibfield  {journal} {\bibinfo  {journal}
  {Phys.~Rev.}\ }\textbf {\bibinfo {volume} {114}},\ \bibinfo {pages} {1219}
  (\bibinfo {year} {1959})}\BibitemShut {NoStop}%
\bibitem [{\citenamefont {Fletcher}\ \emph {et~al.}(1954)\citenamefont
  {Fletcher}, \citenamefont {Yager}, \citenamefont {Pearson},\ and\
  \citenamefont {Merritt}}]{fletcher1954}%
  \BibitemOpen
  \bibfield  {author} {\bibinfo {author} {\bibfnamefont {R.~C.}\ \bibnamefont
  {Fletcher}}, \bibinfo {author} {\bibfnamefont {W.~A.}\ \bibnamefont {Yager}},
  \bibinfo {author} {\bibfnamefont {G.~L.}\ \bibnamefont {Pearson}}, \ and\
  \bibinfo {author} {\bibfnamefont {F.~R.}\ \bibnamefont {Merritt}},\
  }\href@noop {} {\bibfield  {journal} {\bibinfo  {journal} {Phys.~Rev.}\
  }\textbf {\bibinfo {volume} {95}},\ \bibinfo {pages} {844} (\bibinfo {year}
  {1954})}\BibitemShut {NoStop}%
\bibitem [{\citenamefont {Stegner}\ \emph {et~al.}(2010)\citenamefont
  {Stegner}, \citenamefont {Tezuka}, \citenamefont {Andlauer}, \citenamefont
  {Stutzmann}, \citenamefont {Thewalt}, \citenamefont {Brandt},\ and\
  \citenamefont {Itoh}}]{stegner2010}%
  \BibitemOpen
  \bibfield  {author} {\bibinfo {author} {\bibfnamefont {A.~R.}\ \bibnamefont
  {Stegner}}, \bibinfo {author} {\bibfnamefont {H.}~\bibnamefont {Tezuka}},
  \bibinfo {author} {\bibfnamefont {T.}~\bibnamefont {Andlauer}}, \bibinfo
  {author} {\bibfnamefont {M.}~\bibnamefont {Stutzmann}}, \bibinfo {author}
  {\bibfnamefont {M.~L.~W.}\ \bibnamefont {Thewalt}}, \bibinfo {author}
  {\bibfnamefont {M.~S.}\ \bibnamefont {Brandt}}, \ and\ \bibinfo {author}
  {\bibfnamefont {K.~M.}\ \bibnamefont {Itoh}},\ }\href@noop {} {\bibfield
  {journal} {\bibinfo  {journal} {\prb}\ }\textbf {\bibinfo {volume} {82}},\
  \bibinfo {pages} {115213} (\bibinfo {year} {2010})}\BibitemShut {NoStop}%
\bibitem [{\citenamefont {Dirksen}\ \emph {et~al.}(1989)\citenamefont
  {Dirksen}, \citenamefont {Henstra},\ and\ \citenamefont
  {Wenckebach}}]{dirksen1989esr}%
  \BibitemOpen
  \bibfield  {author} {\bibinfo {author} {\bibfnamefont {P.}~\bibnamefont
  {Dirksen}}, \bibinfo {author} {\bibfnamefont {A.}~\bibnamefont {Henstra}}, \
  and\ \bibinfo {author} {\bibfnamefont {W.~T.}\ \bibnamefont {Wenckebach}},\
  }\href@noop {} {\bibfield  {journal} {\bibinfo  {journal}
  {J.~Phys.~Condens.~Matter}\ }\textbf {\bibinfo {volume} {1}},\ \bibinfo
  {pages} {8535} (\bibinfo {year} {1989})}\BibitemShut {NoStop}%
\bibitem [{\citenamefont {Muller}\ \emph {et~al.}(1993)\citenamefont {Muller},
  \citenamefont {Feege}, \citenamefont {Greulich-Weber},\ and\ \citenamefont
  {Spaeth}}]{muller1993endor}%
  \BibitemOpen
  \bibfield  {author} {\bibinfo {author} {\bibfnamefont {R.}~\bibnamefont
  {Muller}}, \bibinfo {author} {\bibfnamefont {M.}~\bibnamefont {Feege}},
  \bibinfo {author} {\bibfnamefont {S.}~\bibnamefont {Greulich-Weber}}, \ and\
  \bibinfo {author} {\bibfnamefont {J.-M.}\ \bibnamefont {Spaeth}},\
  }\href@noop {} {\bibfield  {journal} {\bibinfo  {journal}
  {Semicond.~Sci.~Technol.}\ }\textbf {\bibinfo {volume} {8}},\ \bibinfo
  {pages} {1377} (\bibinfo {year} {1993})}\BibitemShut {NoStop}%
\bibitem [{\citenamefont {Greulich-Weber}\ \emph {et~al.}(1998)\citenamefont
  {Greulich-Weber}, \citenamefont {Feege}, \citenamefont {Kalabukhova},
  \citenamefont {Lukin}, \citenamefont {Spaeth},\ and\ \citenamefont
  {Adrian}}]{greulich1998epr}%
  \BibitemOpen
  \bibfield  {author} {\bibinfo {author} {\bibfnamefont {S.}~\bibnamefont
  {Greulich-Weber}}, \bibinfo {author} {\bibfnamefont {F.}~\bibnamefont
  {Feege}}, \bibinfo {author} {\bibfnamefont {K.~N.}\ \bibnamefont
  {Kalabukhova}}, \bibinfo {author} {\bibfnamefont {S.~N.}\ \bibnamefont
  {Lukin}}, \bibinfo {author} {\bibfnamefont {J.-M.}\ \bibnamefont {Spaeth}}, \
  and\ \bibinfo {author} {\bibfnamefont {F.~J.}\ \bibnamefont {Adrian}},\
  }\href@noop {} {\bibfield  {journal} {\bibinfo  {journal}
  {Semicond.~Sci.~Technol.}\ }\textbf {\bibinfo {volume} {13}},\ \bibinfo
  {pages} {59} (\bibinfo {year} {1998})}\BibitemShut {NoStop}%
\bibitem [{\citenamefont {Rowan}\ \emph {et~al.}(1965)\citenamefont {Rowan},
  \citenamefont {Hahn},\ and\ \citenamefont {Mims}}]{rowan1965}%
  \BibitemOpen
  \bibfield  {author} {\bibinfo {author} {\bibfnamefont {L.~G.}\ \bibnamefont
  {Rowan}}, \bibinfo {author} {\bibfnamefont {E.~L.}\ \bibnamefont {Hahn}}, \
  and\ \bibinfo {author} {\bibfnamefont {W.~B.}\ \bibnamefont {Mims}},\
  }\href@noop {} {\bibfield  {journal} {\bibinfo  {journal} {Phys.~Rev.}\
  }\textbf {\bibinfo {volume} {137}},\ \bibinfo {pages} {A61} (\bibinfo {year}
  {1965})}\BibitemShut {NoStop}%
\bibitem [{\citenamefont {Mims}(1972)}]{mims1972envelope}%
  \BibitemOpen
  \bibfield  {author} {\bibinfo {author} {\bibfnamefont {W.~B.}\ \bibnamefont
  {Mims}},\ }\href@noop {} {\bibfield  {journal} {\bibinfo  {journal} {\prb}\
  }\textbf {\bibinfo {volume} {5}},\ \bibinfo {pages} {2409} (\bibinfo {year}
  {1972})}\BibitemShut {NoStop}%
\bibitem [{\citenamefont {Abe}\ \emph {et~al.}(2004)\citenamefont {Abe},
  \citenamefont {Itoh}, \citenamefont {Isoya},\ and\ \citenamefont
  {Yamasaki}}]{abe2004}%
  \BibitemOpen
  \bibfield  {author} {\bibinfo {author} {\bibfnamefont {E.}~\bibnamefont
  {Abe}}, \bibinfo {author} {\bibfnamefont {K.~M.}\ \bibnamefont {Itoh}},
  \bibinfo {author} {\bibfnamefont {J.}~\bibnamefont {Isoya}}, \ and\ \bibinfo
  {author} {\bibfnamefont {S.}~\bibnamefont {Yamasaki}},\ }\href@noop {}
  {\bibfield  {journal} {\bibinfo  {journal} {\prb}\ }\textbf {\bibinfo
  {volume} {70}},\ \bibinfo {pages} {033204} (\bibinfo {year}
  {2004})}\BibitemShut {NoStop}%
\bibitem [{\citenamefont {Smeltzer}\ \emph {et~al.}(2011)\citenamefont
  {Smeltzer}, \citenamefont {Childress},\ and\ \citenamefont
  {Gali}}]{smeltzer2011}%
  \BibitemOpen
  \bibfield  {author} {\bibinfo {author} {\bibfnamefont {B.}~\bibnamefont
  {Smeltzer}}, \bibinfo {author} {\bibfnamefont {L.}~\bibnamefont {Childress}},
  \ and\ \bibinfo {author} {\bibfnamefont {A.}~\bibnamefont {Gali}},\
  }\href@noop {} {\bibfield  {journal} {\bibinfo  {journal} {New J.~Phys.}\
  }\textbf {\bibinfo {volume} {13}},\ \bibinfo {pages} {025021} (\bibinfo
  {year} {2011})}\BibitemShut {NoStop}%
\bibitem [{\citenamefont {Saikin}\ and\ \citenamefont
  {Fedichkin}(2003)}]{saikin2003}%
  \BibitemOpen
  \bibfield  {author} {\bibinfo {author} {\bibfnamefont {S.}~\bibnamefont
  {Saikin}}\ and\ \bibinfo {author} {\bibfnamefont {L.}~\bibnamefont
  {Fedichkin}},\ }\href@noop {} {\bibfield  {journal} {\bibinfo  {journal}
  {\prb}\ }\textbf {\bibinfo {volume} {67}},\ \bibinfo {pages} {161302(R)}
  (\bibinfo {year} {2003})}\BibitemShut {NoStop}%
\bibitem [{\citenamefont {Witzel}\ \emph {et~al.}(2007)\citenamefont {Witzel},
  \citenamefont {Hu},\ and\ \citenamefont {{Das Sarma}}}]{witzel2007}%
  \BibitemOpen
  \bibfield  {author} {\bibinfo {author} {\bibfnamefont {W.~M.}\ \bibnamefont
  {Witzel}}, \bibinfo {author} {\bibfnamefont {X.}~\bibnamefont {Hu}}, \ and\
  \bibinfo {author} {\bibfnamefont {S.}~\bibnamefont {{Das Sarma}}},\
  }\href@noop {} {\bibfield  {journal} {\bibinfo  {journal} {\prb}\ }\textbf
  {\bibinfo {volume} {76}},\ \bibinfo {pages} {035212} (\bibinfo {year}
  {2007})}\BibitemShut {NoStop}%
\bibitem [{\citenamefont {Maze}\ \emph {et~al.}(2008)\citenamefont {Maze},
  \citenamefont {Taylor},\ and\ \citenamefont {Lukin}}]{maze2008}%
  \BibitemOpen
  \bibfield  {author} {\bibinfo {author} {\bibfnamefont {J.~R.}\ \bibnamefont
  {Maze}}, \bibinfo {author} {\bibfnamefont {J.~M.}\ \bibnamefont {Taylor}}, \
  and\ \bibinfo {author} {\bibfnamefont {M.~D.}\ \bibnamefont {Lukin}},\
  }\href@noop {} {\bibfield  {journal} {\bibinfo  {journal} {\prb}\ }\textbf
  {\bibinfo {volume} {78}},\ \bibinfo {pages} {094303} (\bibinfo {year}
  {2008})}\BibitemShut {NoStop}%
\bibitem [{\citenamefont {Wang}\ \emph {et~al.}(2012)\citenamefont {Wang},
  \citenamefont {Chesi},\ and\ \citenamefont {Coish}}]{wang2012}%
  \BibitemOpen
  \bibfield  {author} {\bibinfo {author} {\bibfnamefont {X.~J.}\ \bibnamefont
  {Wang}}, \bibinfo {author} {\bibfnamefont {S.}~\bibnamefont {Chesi}}, \ and\
  \bibinfo {author} {\bibfnamefont {W.~A.}\ \bibnamefont {Coish}},\ }\href@noop
  {} {\bibfield  {journal} {\bibinfo  {journal} {\prl}\ }\textbf {\bibinfo
  {volume} {109}},\ \bibinfo {pages} {237601} (\bibinfo {year}
  {2012})}\BibitemShut {NoStop}%
\bibitem [{\citenamefont {Wang}\ \emph {et~al.}(2015)\citenamefont {Wang},
  \citenamefont {Chesi},\ and\ \citenamefont {Coish}}]{wang2015}%
  \BibitemOpen
  \bibfield  {author} {\bibinfo {author} {\bibfnamefont {X.~J.}\ \bibnamefont
  {Wang}}, \bibinfo {author} {\bibfnamefont {S.}~\bibnamefont {Chesi}}, \ and\
  \bibinfo {author} {\bibfnamefont {W.~A.}\ \bibnamefont {Coish}},\ }\href@noop
  {} {\bibfield  {journal} {\bibinfo  {journal} {\prb}\ }\textbf {\bibinfo
  {volume} {92}},\ \bibinfo {pages} {115424} (\bibinfo {year}
  {2015})}\BibitemShut {NoStop}%
\bibitem [{\citenamefont {Carter}\ \emph {et~al.}(2014)\citenamefont {Carter},
  \citenamefont {Economou}, \citenamefont {Greilich}, \citenamefont {Barnes},
  \citenamefont {Sweeney}, \citenamefont {Bracker},\ and\ \citenamefont
  {Gammon}}]{carter2014}%
  \BibitemOpen
  \bibfield  {author} {\bibinfo {author} {\bibfnamefont {S.~G.}\ \bibnamefont
  {Carter}}, \bibinfo {author} {\bibfnamefont {S.~E.}\ \bibnamefont
  {Economou}}, \bibinfo {author} {\bibfnamefont {A.}~\bibnamefont {Greilich}},
  \bibinfo {author} {\bibfnamefont {E.}~\bibnamefont {Barnes}}, \bibinfo
  {author} {\bibfnamefont {T.}~\bibnamefont {Sweeney}}, \bibinfo {author}
  {\bibfnamefont {A.~S.}\ \bibnamefont {Bracker}}, \ and\ \bibinfo {author}
  {\bibfnamefont {D.}~\bibnamefont {Gammon}},\ }\href@noop {} {\bibfield
  {journal} {\bibinfo  {journal} {\prb}\ }\textbf {\bibinfo {volume} {89}},\
  \bibinfo {pages} {075316} (\bibinfo {year} {2014})}\BibitemShut {NoStop}%
\bibitem [{\citenamefont {Slichter}(1978)}]{slichter1978}%
  \BibitemOpen
  \bibfield  {author} {\bibinfo {author} {\bibfnamefont {C.~P.}\ \bibnamefont
  {Slichter}},\ }\href@noop {} {\emph {\bibinfo {title} {Principles of magnetic
  resonance}}},\ \bibinfo {edition} {2nd}\ ed.\ (\bibinfo  {publisher}
  {Springer},\ \bibinfo {year} {1978})\BibitemShut {NoStop}%
\bibitem [{\citenamefont {Philippopoulos}\ \emph {et~al.}(tion)\citenamefont
  {Philippopoulos}, \citenamefont {Chesi},\ and\ \citenamefont
  {Coish}}]{philippopoulos2019}%
  \BibitemOpen
  \bibfield  {author} {\bibinfo {author} {\bibfnamefont {P.}~\bibnamefont
  {Philippopoulos}}, \bibinfo {author} {\bibfnamefont {S.}~\bibnamefont
  {Chesi}}, \ and\ \bibinfo {author} {\bibfnamefont {W.~A.}\ \bibnamefont
  {Coish}},\ }\href@noop {} {\  (\bibinfo {year} {in preparation})}\BibitemShut
  {NoStop}%
\bibitem [{\citenamefont {Abragam}(1961)}]{abragam1961principles}%
  \BibitemOpen
  \bibfield  {author} {\bibinfo {author} {\bibfnamefont {A.}~\bibnamefont
  {Abragam}},\ }\href@noop {} {\emph {\bibinfo {title} {The principles of
  nuclear magnetism}}},\ \bibinfo {number} {32}\ (\bibinfo  {publisher} {Oxford
  university press},\ \bibinfo {year} {1961})\BibitemShut {NoStop}%
\bibitem [{\citenamefont {Mims}\ \emph {et~al.}(1977)\citenamefont {Mims},
  \citenamefont {Peisach},\ and\ \citenamefont {Davis}}]{mims1977nuclear}%
  \BibitemOpen
  \bibfield  {author} {\bibinfo {author} {\bibfnamefont {W.~B.}\ \bibnamefont
  {Mims}}, \bibinfo {author} {\bibfnamefont {J.}~\bibnamefont {Peisach}}, \
  and\ \bibinfo {author} {\bibfnamefont {J.~L.}\ \bibnamefont {Davis}},\
  }\href@noop {} {\bibfield  {journal} {\bibinfo  {journal} {\jcp}\ }\textbf
  {\bibinfo {volume} {66}},\ \bibinfo {pages} {5536} (\bibinfo {year}
  {1977})}\BibitemShut {NoStop}%
\bibitem [{\citenamefont {Debernardi}\ \emph {et~al.}(2006)\citenamefont
  {Debernardi}, \citenamefont {Baldereschi},\ and\ \citenamefont
  {Fanciulli}}]{debernardi2006}%
  \BibitemOpen
  \bibfield  {author} {\bibinfo {author} {\bibfnamefont {A.}~\bibnamefont
  {Debernardi}}, \bibinfo {author} {\bibfnamefont {A.}~\bibnamefont
  {Baldereschi}}, \ and\ \bibinfo {author} {\bibfnamefont {M.}~\bibnamefont
  {Fanciulli}},\ }\href@noop {} {\bibfield  {journal} {\bibinfo  {journal}
  {\prb}\ }\textbf {\bibinfo {volume} {74}},\ \bibinfo {pages} {035202}
  (\bibinfo {year} {2006})}\BibitemShut {NoStop}%
\bibitem [{\citenamefont {Usman}\ \emph {et~al.}(2015)\citenamefont {Usman},
  \citenamefont {Rahman}, \citenamefont {Salfi}, \citenamefont {Bocquel},
  \citenamefont {Voisin}, \citenamefont {Rogge}, \citenamefont {Klimeck},\ and\
  \citenamefont {Hollenberg}}]{usman2015}%
  \BibitemOpen
  \bibfield  {author} {\bibinfo {author} {\bibfnamefont {M.}~\bibnamefont
  {Usman}}, \bibinfo {author} {\bibfnamefont {R.}~\bibnamefont {Rahman}},
  \bibinfo {author} {\bibfnamefont {J.}~\bibnamefont {Salfi}}, \bibinfo
  {author} {\bibfnamefont {J.}~\bibnamefont {Bocquel}}, \bibinfo {author}
  {\bibfnamefont {B.}~\bibnamefont {Voisin}}, \bibinfo {author} {\bibfnamefont
  {S.}~\bibnamefont {Rogge}}, \bibinfo {author} {\bibfnamefont
  {G.}~\bibnamefont {Klimeck}}, \ and\ \bibinfo {author} {\bibfnamefont
  {L.~L.~C.}\ \bibnamefont {Hollenberg}},\ }\href@noop {} {\bibfield  {journal}
  {\bibinfo  {journal} {J.~Phys.~Condens.~Matter}\ }\textbf {\bibinfo {volume}
  {27}},\ \bibinfo {pages} {154207} (\bibinfo {year} {2015})}\BibitemShut
  {NoStop}%
\bibitem [{\citenamefont {Pica}\ \emph {et~al.}(2014)\citenamefont {Pica},
  \citenamefont {Wolfowicz}, \citenamefont {Urdampilleta}, \citenamefont
  {Thewalt}, \citenamefont {Riemann}, \citenamefont {Abrosimov}, \citenamefont
  {Becker}, \citenamefont {Pohl}, \citenamefont {Morton}, \citenamefont {Bhatt}
  \emph {et~al.}}]{pica2014}%
  \BibitemOpen
  \bibfield  {author} {\bibinfo {author} {\bibfnamefont {G.}~\bibnamefont
  {Pica}}, \bibinfo {author} {\bibfnamefont {G.}~\bibnamefont {Wolfowicz}},
  \bibinfo {author} {\bibfnamefont {M.}~\bibnamefont {Urdampilleta}}, \bibinfo
  {author} {\bibfnamefont {M.~L.~W.}\ \bibnamefont {Thewalt}}, \bibinfo
  {author} {\bibfnamefont {H.}~\bibnamefont {Riemann}}, \bibinfo {author}
  {\bibfnamefont {N.~V.}\ \bibnamefont {Abrosimov}}, \bibinfo {author}
  {\bibfnamefont {P.}~\bibnamefont {Becker}}, \bibinfo {author} {\bibfnamefont
  {H.-J.}\ \bibnamefont {Pohl}}, \bibinfo {author} {\bibfnamefont {J.~J.~L.}\
  \bibnamefont {Morton}}, \bibinfo {author} {\bibfnamefont {R.~N.}\
  \bibnamefont {Bhatt}},  \emph {et~al.},\ }\href@noop {} {\bibfield  {journal}
  {\bibinfo  {journal} {\prb}\ }\textbf {\bibinfo {volume} {90}},\ \bibinfo
  {pages} {195204} (\bibinfo {year} {2014})}\BibitemShut {NoStop}%
\bibitem [{\citenamefont {Kohn}\ and\ \citenamefont
  {Luttinger}(1955{\natexlab{a}})}]{kohn1955hyperfine}%
  \BibitemOpen
  \bibfield  {author} {\bibinfo {author} {\bibfnamefont {W.}~\bibnamefont
  {Kohn}}\ and\ \bibinfo {author} {\bibfnamefont {J.~M.}\ \bibnamefont
  {Luttinger}},\ }\href@noop {} {\bibfield  {journal} {\bibinfo  {journal}
  {Phys.~Rev.}\ }\textbf {\bibinfo {volume} {97}},\ \bibinfo {pages} {883}
  (\bibinfo {year} {1955}{\natexlab{a}})}\BibitemShut {NoStop}%
\bibitem [{Note1()}]{Note1}%
  \BibitemOpen
  \bibinfo {note} {Here, we use the term `electrostatics' to describe the
  problem of finding the full wavefunction of the impurity-bound hole. This
  includes the long-range properties (band properties like the effective mass)
  and short-range properties (such as the core electronic structure of the
  impurity atom). If the electrostatics of the phosphorus donor and the boron
  acceptor are similar, then the rescaling parameter (central-cell corrections)
  should also be similar. While the similarity in binding energies is fully
  consistent with the two impurities having similar electrostatics, we cannot
  rule out the possibility of an accidental coincidence in binding energy.
  Ultimately, the true value of the hyperfine coupling (and central-cell
  correction) should be established experimentally, as proposed
  here.}\BibitemShut {Stop}%
\bibitem [{\citenamefont {Kohn}\ and\ \citenamefont
  {Luttinger}(1955{\natexlab{b}})}]{kohn1955theory}%
  \BibitemOpen
  \bibfield  {author} {\bibinfo {author} {\bibfnamefont {W.}~\bibnamefont
  {Kohn}}\ and\ \bibinfo {author} {\bibfnamefont {J.~M.}\ \bibnamefont
  {Luttinger}},\ }\href@noop {} {\bibfield  {journal} {\bibinfo  {journal}
  {Phys.~Rev.}\ }\textbf {\bibinfo {volume} {98}},\ \bibinfo {pages} {915}
  (\bibinfo {year} {1955}{\natexlab{b}})}\BibitemShut {NoStop}%
\bibitem [{\citenamefont {Coish}\ and\ \citenamefont
  {Baugh}(2009)}]{coish2009nuclear}%
  \BibitemOpen
  \bibfield  {author} {\bibinfo {author} {\bibfnamefont {W.~A.}\ \bibnamefont
  {Coish}}\ and\ \bibinfo {author} {\bibfnamefont {J.}~\bibnamefont {Baugh}},\
  }\href@noop {} {\bibfield  {journal} {\bibinfo  {journal}
  {Phys.~Status~Solidi~B}\ }\textbf {\bibinfo {volume} {246}},\ \bibinfo
  {pages} {2203} (\bibinfo {year} {2009})}\BibitemShut {NoStop}%
\bibitem [{\citenamefont {Ohkawa}(1979)}]{ohkawa1979}%
  \BibitemOpen
  \bibfield  {author} {\bibinfo {author} {\bibfnamefont {F.~J.}\ \bibnamefont
  {Ohkawa}},\ }\href@noop {} {\bibfield  {journal} {\bibinfo  {journal}
  {J.~Phys.~Soc.~Jpn.}\ }\textbf {\bibinfo {volume} {46}},\ \bibinfo {pages}
  {1529} (\bibinfo {year} {1979})}\BibitemShut {NoStop}%
\bibitem [{\citenamefont {Cardona}\ and\ \citenamefont
  {Pollak}(1966)}]{cardona1966}%
  \BibitemOpen
  \bibfield  {author} {\bibinfo {author} {\bibfnamefont {M.}~\bibnamefont
  {Cardona}}\ and\ \bibinfo {author} {\bibfnamefont {F.~H.}\ \bibnamefont
  {Pollak}},\ }\href@noop {} {\bibfield  {journal} {\bibinfo  {journal}
  {Phys.~Rev.}\ }\textbf {\bibinfo {volume} {142}},\ \bibinfo {pages} {530}
  (\bibinfo {year} {1966})}\BibitemShut {NoStop}%
\bibitem [{\citenamefont {Chekhovich}\ \emph {et~al.}(2013)\citenamefont
  {Chekhovich}, \citenamefont {Glazov}, \citenamefont {Krysa}, \citenamefont
  {Hopkinson}, \citenamefont {Senellart}, \citenamefont {Lemaitre},
  \citenamefont {Skolnick},\ and\ \citenamefont
  {Tartakovskii}}]{chekhovich2013element}%
  \BibitemOpen
  \bibfield  {author} {\bibinfo {author} {\bibfnamefont {E.~A.}\ \bibnamefont
  {Chekhovich}}, \bibinfo {author} {\bibfnamefont {M.~M.}\ \bibnamefont
  {Glazov}}, \bibinfo {author} {\bibfnamefont {A.~B.}\ \bibnamefont {Krysa}},
  \bibinfo {author} {\bibfnamefont {M.}~\bibnamefont {Hopkinson}}, \bibinfo
  {author} {\bibfnamefont {P.}~\bibnamefont {Senellart}}, \bibinfo {author}
  {\bibfnamefont {A.}~\bibnamefont {Lemaitre}}, \bibinfo {author}
  {\bibfnamefont {M.~S.}\ \bibnamefont {Skolnick}}, \ and\ \bibinfo {author}
  {\bibfnamefont {A.~I.}\ \bibnamefont {Tartakovskii}},\ }\href@noop {}
  {\bibfield  {journal} {\bibinfo  {journal} {Nat.~Phys.}\ }\textbf {\bibinfo
  {volume} {9}},\ \bibinfo {pages} {74} (\bibinfo {year} {2013})}\BibitemShut
  {NoStop}%
\bibitem [{Note2()}]{Note2}%
  \BibitemOpen
  \bibinfo {note} {In the case of a flat (quasi-2D) unstrained quantum dot, the
  effective mass results in a heavy-hole ($J_3=\pm 3/2$) ground state. For an
  acceptor impurity or quantum dot under biaxial in-plane tensile strain along
  $[100]$ and $[010]$, the light-hole ($J_3=\pm 1/2$) states may describe the
  ground state. The $x_3$ axis is the out-of-plane direction (e.g., the growth
  axis for a quantum dot defined in a 2D hole gas at a heterostructure
  interface). For example, for a growth axis along $[001]$, we could take $x_3
  = [001]$, $x_1 =[100]$, and $x_2=[010]$.}\BibitemShut {Stop}%
\bibitem [{\citenamefont {Yafet}(1961)}]{yafet1961}%
  \BibitemOpen
  \bibfield  {author} {\bibinfo {author} {\bibfnamefont {Y.}~\bibnamefont
  {Yafet}},\ }\href@noop {} {\bibfield  {journal} {\bibinfo  {journal}
  {J.~Phys.~Chem.~Solids}\ }\textbf {\bibinfo {volume} {21}},\ \bibinfo {pages}
  {99} (\bibinfo {year} {1961})}\BibitemShut {NoStop}%
\bibitem [{\citenamefont {Clementi}\ and\ \citenamefont
  {Raimondi}(1963)}]{clementi1963}%
  \BibitemOpen
  \bibfield  {author} {\bibinfo {author} {\bibfnamefont {E.}~\bibnamefont
  {Clementi}}\ and\ \bibinfo {author} {\bibfnamefont {D.~L.}\ \bibnamefont
  {Raimondi}},\ }\href@noop {} {\bibfield  {journal} {\bibinfo  {journal}
  {\jcp}\ }\textbf {\bibinfo {volume} {38}},\ \bibinfo {pages} {2686} (\bibinfo
  {year} {1963})}\BibitemShut {NoStop}%
\bibitem [{\citenamefont {Burstein}\ \emph {et~al.}(1956)\citenamefont
  {Burstein}, \citenamefont {Picus}, \citenamefont {Henvis},\ and\
  \citenamefont {Wallis}}]{burstein1956}%
  \BibitemOpen
  \bibfield  {author} {\bibinfo {author} {\bibfnamefont {E.}~\bibnamefont
  {Burstein}}, \bibinfo {author} {\bibfnamefont {G.}~\bibnamefont {Picus}},
  \bibinfo {author} {\bibfnamefont {B.}~\bibnamefont {Henvis}}, \ and\ \bibinfo
  {author} {\bibfnamefont {R.}~\bibnamefont {Wallis}},\ }\href@noop {}
  {\bibfield  {journal} {\bibinfo  {journal} {J.~Phys.~Chem.~Solids}\ }\textbf
  {\bibinfo {volume} {1}},\ \bibinfo {pages} {65} (\bibinfo {year}
  {1956})}\BibitemShut {NoStop}%
\bibitem [{\citenamefont {Bir}\ \emph {et~al.}(1963{\natexlab{a}})\citenamefont
  {Bir}, \citenamefont {Butekov},\ and\ \citenamefont {Pikus}}]{bir1963spin}%
  \BibitemOpen
  \bibfield  {author} {\bibinfo {author} {\bibfnamefont {G.~L.}\ \bibnamefont
  {Bir}}, \bibinfo {author} {\bibfnamefont {E.~I.}\ \bibnamefont {Butekov}}, \
  and\ \bibinfo {author} {\bibfnamefont {G.~E.}\ \bibnamefont {Pikus}},\
  }\href@noop {} {\bibfield  {journal} {\bibinfo  {journal}
  {J.~Phys.~Chem.~Solids}\ }\textbf {\bibinfo {volume} {24}},\ \bibinfo {pages}
  {1467} (\bibinfo {year} {1963}{\natexlab{a}})}\BibitemShut {NoStop}%
\bibitem [{\citenamefont {Bir}\ \emph {et~al.}(1963{\natexlab{b}})\citenamefont
  {Bir}, \citenamefont {Butikov},\ and\ \citenamefont {Pikus}}]{bir1963spinII}%
  \BibitemOpen
  \bibfield  {author} {\bibinfo {author} {\bibfnamefont {G.~L.}\ \bibnamefont
  {Bir}}, \bibinfo {author} {\bibfnamefont {E.~I.}\ \bibnamefont {Butikov}}, \
  and\ \bibinfo {author} {\bibfnamefont {G.~E.}\ \bibnamefont {Pikus}},\
  }\href@noop {} {\bibfield  {journal} {\bibinfo  {journal}
  {J.~Phys.~Chem.~Solids}\ }\textbf {\bibinfo {volume} {24}},\ \bibinfo {pages}
  {1475} (\bibinfo {year} {1963}{\natexlab{b}})}\BibitemShut {NoStop}%
\bibitem [{\citenamefont {Overhof}\ and\ \citenamefont
  {Gerstmann}(2004)}]{overhof2004ab}%
  \BibitemOpen
  \bibfield  {author} {\bibinfo {author} {\bibfnamefont {H.}~\bibnamefont
  {Overhof}}\ and\ \bibinfo {author} {\bibfnamefont {U.}~\bibnamefont
  {Gerstmann}},\ }\href@noop {} {\bibfield  {journal} {\bibinfo  {journal}
  {\prl}\ }\textbf {\bibinfo {volume} {92}},\ \bibinfo {pages} {087602}
  (\bibinfo {year} {2004})}\BibitemShut {NoStop}%
\bibitem [{\citenamefont {Hodges}\ \emph {et~al.}(2008)\citenamefont {Hodges},
  \citenamefont {Yang}, \citenamefont {Ramanathan},\ and\ \citenamefont
  {Cory}}]{hodges2008}%
  \BibitemOpen
  \bibfield  {author} {\bibinfo {author} {\bibfnamefont {J.~S.}\ \bibnamefont
  {Hodges}}, \bibinfo {author} {\bibfnamefont {J.~C.}\ \bibnamefont {Yang}},
  \bibinfo {author} {\bibfnamefont {C.}~\bibnamefont {Ramanathan}}, \ and\
  \bibinfo {author} {\bibfnamefont {D.~G.}\ \bibnamefont {Cory}},\ }\href@noop
  {} {\bibfield  {journal} {\bibinfo  {journal} {\pra}\ }\textbf {\bibinfo
  {volume} {78}},\ \bibinfo {pages} {010303(R)} (\bibinfo {year}
  {2008})}\BibitemShut {NoStop}%
\bibitem [{\citenamefont {Luttinger}(1956)}]{luttinger1956}%
  \BibitemOpen
  \bibfield  {author} {\bibinfo {author} {\bibfnamefont {J.~M.}\ \bibnamefont
  {Luttinger}},\ }\href@noop {} {\bibfield  {journal} {\bibinfo  {journal}
  {Phys.~Rev.}\ }\textbf {\bibinfo {volume} {102}},\ \bibinfo {pages} {1030}
  (\bibinfo {year} {1956})}\BibitemShut {NoStop}%
\bibitem [{\citenamefont {Winkler}(2003)}]{winkler2003}%
  \BibitemOpen
  \bibfield  {author} {\bibinfo {author} {\bibfnamefont {R.}~\bibnamefont
  {Winkler}},\ }\href@noop {} {\emph {\bibinfo {title} {Spin-Orbit Coupling
  Effects in Two-Dimensional Electron and Hole Systems}}}\ (\bibinfo
  {publisher} {Springer},\ \bibinfo {year} {2003})\BibitemShut {NoStop}%
\bibitem [{Note3()}]{Note3}%
  \BibitemOpen
  \bibinfo {note} {The convention used here is that the pseudo-angular-momentum
  $\protect \bf {J}$ measures the pseudo-angular-momentum of the crystal.
  Therefore, a hole in state $J_3 = m$ indicates that an electron with $J_3 =
  -m$ has been anihilated from the filled valence band, and the remaining
  member of the Kramers doublet has $J_3 = m$.}\BibitemShut {Stop}%
\end{thebibliography}%

\end{document}